\documentclass[aps,pre,superscriptaddress,reprint]{revtex4-2}
\usepackage{hyperref} 
\usepackage{verbatim} 
\usepackage{epsfig,amsmath,amssymb,calc,verbatim}
\usepackage{bm}
\usepackage{subcaption}
\usepackage{ragged2e}
\usepackage{physics}

\renewcommand{\phi}{\varphi}
\newcommand{\phiJ}{\varphi_{\mathrm{J}}}
\newcommand{\Diameter}{D}
\newcommand{\DiameterS}{D_{\mathrm{S}}}
\newcommand{\DiameterL}{D_{\mathrm{L}}}

\newcommand{\calO}{\mathcal{O}}
\newcommand{\gammaz}{\gamma}
\newcommand{\gammas}{\gamma_{\mathrm{s}}}

\newcommand{\gammay}{\gamma_{\mathrm{y}}}
\newcommand{\omegas}{\omega_{\mathrm{s}}}
\newcommand{\sigmay}{\sigma_{\mathrm{y}}}
\newcommand{\GAQS}{G_{\mathrm{qs}}}
\newcommand{\Dr}{{\delta r}}
\newcommand{\dphi}{\delta\varphi}

\begin{document}

\preprint{APS/123-QED}

\title{Pre-yielding mechanical response near the jamming transition}
\author{Hidemasa Bessho}
\email{bessho@r.phys.nagoya-u.ac.jp}
\affiliation{Department of Physics, Nagoya University, Nagoya 464-8602, Japan}
\author{Takeshi Kawasaki}
\affiliation{Department of Physics, Nagoya University, Nagoya 464-8602, Japan}
\affiliation{D3 Center, The University of Osaka, Toyonaka, Osaka 560-0043, Japan}
\affiliation{Department of Physics, The University of Osaka, Toyonaka, Osaka 560-0043, Japan}
\author{Kunimasa Miyazaki}
\email{miyazaki@r.phys.nagoya-u.ac.jp}
\affiliation{Department of Physics, Nagoya University, Nagoya 464-8602, Japan}

\begin{abstract}
The mechanical and rheological properties of jammed packings of frictionless particles 
under shear strain remain not fully understood, even when the strain amplitude is very 
small and well below the yielding threshold. 
Systems above the jamming transition point $\phiJ$ are known to display two
     anomalous mechanical behaviors with respect to the driving frequency $\omega$ (or
     time $t$) and the strain amplitude $\gamma$. 
In the linear-response regime ($\gamma\to 0$),  
the complex modulus exhibits an algebraic scaling, 
$G(\omega)\sim\omega^{1/2}$ (or  $G(t)\sim t^{-1/2}$ in the time representation). 
In contrast, in the quasi-static limit ($\omega \to 0$), 
the modulus shows the nonlinear behavior, $G(\gamma)\sim\gamma^{-1/2}$, a phenomenon 
referred to as softening.  
The ranges of $\omega$ and $\gamma$ over which these algebraic scalings hold broaden 
as $\phiJ$ is approached from above, whereas both $G(\omega)$ and $G(\gamma)$
     vanish for $\phi < \phiJ$.    
In this study, we investigate the mechanical response 
in the regime where these two anomalies coexist in the vicinity of $\phiJ$.
To this end, we perform numerical analyses using two rheological protocols:
     oscillatory shear and transient stress relaxation.
Our results demonstrate that the mechanical responses are not simply described as a superposition
of the two algebraic relaxations and instead exhibit rich nonlinear viscoelastic behavior 
both above and even below $\phiJ$.    
\end{abstract}

\maketitle

\section{Introduction}
Disordered packings of athermal particles, such as emulsions, non-Brownian
suspensions, granular materials, and foams, become rigid when their density, or
packing fraction $\phi$, exceeds the jamming transition point $\phiJ$~\cite{Liu_Nagel}.  
For frictionless spherical particles, various scaling laws and universal behaviors near $\phiJ$ 
have been established~\cite{O'Hern2003pre,vanHecke2010}. 
In particular, the algebraic dependence of the contact number and pressure on
$\dphi=\phi-\phiJ$,  the emergence of diverging length scales, and the
universal spectrum of the vibrational density of states have been identified in
both simulations~\cite{O'Hern2003pre,Silbert2005prl,Mizuno2017pnas} and
experiments~\cite{Majmudar2007prl,Katgert_2010}, and explained by
theoretical approaches~\cite{Wyart2005ap,DeGiuli2014sm,Parisi_Urbani_Zamponi_2020}. 
These results characterize the properties of packings in the absence
of applied stress.

Jammed systems also exhibit rich mechanical and rheological behaviors when 
subjected to external stress or shear deformation.
Thus far, studies of the mechanical response near the jamming transition have primarily followed two main directions: steady-shear rheology and the response to small shear strains.
In the former case, the main focus is on the relationship between shear stress $\sigma$ 
and the shear rate $\dot{\gamma}$. 
For example, 
the shear viscosity, $\eta\equiv \sigma/\dot{\gamma}$, of non-Brownian
suspensions diverges as $\eta \sim |\dphi|^{-\beta}$ below $\phiJ$,
while, above $\phiJ$, the shear stress is well described by the 
Herschel--Bulkley law, $\sigma =\sigmay+C\dot{\gamma}^{\,n}$.   
The positive exponents $\beta$, $n$, together with the yield stress defined as
$\sigmay\equiv\sigma(\dot{\gamma}\to0)$, serve as key indicators of the critical
behavior associated with the jamming transition~\cite{Durian1995prl,Olsson2007prl,Otsuki2009pre,Hatano2008jpsj,Hatano2010ptp,Tighe2010prl,Boyer2011prl,Olsson2011pre,Lerner2012pnas,Dinkgreve2015pre,Kawasaki2015pre,Olsson2015pre,Vagberg2016pre,deGiuli2015pre,Bonn2017rmp,Ikeda2020prl}. 

The latter case, the response to small shear strains, is the main focus of this study. 
A jammed packing behaves as an elastic solid at very small $\gamma$ but 
upon increasing $\gamma$, it yields 
and enters the plastic regime at a finite strain
$\gammay$~\cite{Pan2023pr,Berthier2025natrevphys,Divoux2024sm}.   
It is known that, even in the pre-yielding regime, where $\gamma$ is well below  $\gammay$, 
jammed packings exhibit highly nontrivial mechanical properties, owing to the marginal stability
inherent to jamming criticality, in contrast to other amorphous solids.  
In particular, two distinct anomalous algebraic scalings for the mechanical response
function are observed. 
One is the scale-free frequency dependence of the viscoelastic complex modulus. 
As the system approaches $\phiJ$ from above, the complex shear modulus, 
$G^*(\omega) = G'(\omega)+iG''(\omega)$, 
develops a power-law behavior, $G^*(\omega) \approx  A\omega^{1/2}$ 
(hereafter referred to as the
``$\omega^{1/2}$ scaling'')~\cite{Tighe2011prl,Baumgarten2017sm,Hara2025natp}.  
The $\omega^{1/2}$ scaling is universally observed, regardless of system
details, over a wide range of frequencies.  
This scaling is observed for $\omega > \omegas$, where $\omegas$ is the onset
frequency that scales as $\omegas \propto \dphi$ for the harmonic potentials (and
$\propto \dphi^{\alpha-1}$ for other soft-repulsive potentials defined in
eqn~(\ref{eq:potential})).  
The $\omega^{1/2}$ scaling is understood as a consequence of the
development of the plateau in the vibrational density of states near
$\phiJ$~\cite{vanHecke2010} and the boson peak,  
universally observed in amorphous
solids~\cite{Phillips1981book,Schirmacher2007prl,Charbonneau2016prl,Mizuno2017pnas}.  
A real-time manifestation of the $\omega^{1/2}$ scaling is observed in the
transient stress relaxation after a step strain, where the modulus decays as
$G(t) \equiv \sigma(t)/\gamma \sim t^{-1/2}$~\cite{Hatano,Boschan2,Saitoh2020prl}.
The other anomalous algebraic scaling is the 
nonlinear stress response in the quasi-static limit ($\omega\to 0$), 
known as {\it shear softening}. 
The static storage modulus $\GAQS \equiv G^{\prime}(\omega\rightarrow 0)$, which
is independent of $\gamma$ in the linear-response regime,  
starts to exhibit nonlinear softening behavior, $\GAQS \sim \gamma^{-1/2}$
as $\phiJ$ is approached from
above~\cite{Otsuki2014,Boschan2016sm,Dagois-Bohy2017sm,Otsuki2022prl,Kawasaki2024prl}.  
We refer to the softening as the ``$\gamma^{1/2}$ scaling'', hereafter.
The softening exponent $1/2$ is insensitive to system details.
The onset strain $\gammas$ that marks the crossover from the elastic to the
softening regime is proportional to $\dphi$. 
Recently, we have found a scaling law that unifies the elastic, softening,
and the yielding regimes~\cite{Kawasaki2024prl}. 

In this paper, we investigate the interplay between the $\omega^{1/2}$ 
scaling and $\gamma^{1/2}$ scaling of jammed packings in the pre-yielding regime
both above and below $\phiJ$. 
We address several key questions in this work.  
First, how do the $\omega^{1/2}$ scaling (or $t^{-1/2}$ scaling) and
$\gamma^{1/2}$ scaling coexist?
Previous studies of the $\omega^{1/2}$ scaling assume small $\gamma$,  
whereas studies of the $\gamma^{1/2}$ scaling generally focus on the quasi-static
 limit.
 However, as $\phiJ$ is approached, the two scaling regimes inevitably broaden
 and overlap. 
Therefore, in the vicinity of $\phiJ$, the two scalings become simultaneously relevant, 
and their interplay is important. 
If both scalings for $\omega$ and $\gamma$ coexist, should we expect
a simple superposition of the two scaling functions of $\omega$ and $\gamma$, analogous to
the Cox--Merz rule~\cite{larson1999structure} or the strain-rate frequency superposition~\cite{Wyss2007prl}? 

Second, how does the mechanical response above $\phiJ$ cross over to that below $\phiJ$? 
$G^{\ast}(\omega)$ in the linear-response regime ($\gamma \to 0$)
and $\GAQS(\gamma)$ in the quasi-static limit ($\omega\to 0$) are finite only above $\phiJ$.
Both are strictly zero below $\phiJ$, as a persistent contact network is
absent. 
If $\omega$ and $\gamma$ are finite, the stress may arise even
below $\phiJ$, because particles intermittently collide and form transient contact networks.
The formation of the contact networks of frictionless spheres under oscillatory shear 
below--but close to--$\phiJ$ has been extensively studied
in the context of the reversible-irreversible (or absorbing) transition 
~\cite{Milz2013pre,Schreck2013pre,Nagasawa2019sm,Matsuyama2021epje,Das2010pnas}.     
However, the relationship between this nonequilibrium phase transition and the
nonlinear mechanical response remains largely unexplored. 

Third, how is the stress relaxation of jammed packings influenced by 
nonlinear perturbations? 
The relaxation dynamics of amorphous solids from perturbed configurations 
toward a ground state, \textit{i.e.},  toward the bottom of the potential-energy
landscape, are important for understanding the hierarchical structure 
of rugged energy landscape in the context of the glass
transition~\cite{Chacko2019prl,Folena2020prx,Ikeda2020prl,Nishikawa2022prx}. 
Studying the $\gamma$-dependence of $G(t)$ for various $\gamma$ would thus provide
insight into the connection between the nonlinear rheology and the
topographical features of the energy landscape.

Our study is the first step toward addressing these questions.
We perform numerical simulations and investigate the nonlinear mechanical responses under
time-dependent shear strains $\gamma(t)$ in frictionless, athermal jammed packings
near $\phiJ$ using different shear protocols.
The first protocol is the oscillatory-shear measurement, designed to address the first and second questions.
By fixing the frequency within the $\omega^{1/2}$ scaling regime and varying the
strain amplitude over a wide range, we measure the complex shear modulus
$G^*(\omega)$ both above and below $\phiJ$. 
While many studies have attempted to unify the mechanical responses above and below $\phiJ$ 
in the post-yielding regime, $\gamma \rightarrow \infty$~\cite{Hatano2008jpsj,Olsson2007prl,Hatano2010ptp,Otsuki2009pre,Tighe2010prl,Vagberg2016pre}, 
little attention has been paid to the pre-yielding regime $\gamma < \gammay$. 
The second is transient stress relaxation, in which  
the time-dependence of the stress after a step shear strain is monitored.
It is equivalent to the oscillatory-shear measurement in the linear-response regime, but
deviates from it once the linear-response assumption breaks down.
We analyze the transient relaxation of shear stress, energy, force, and
displacement following instantaneous strain deformation over a broad range of
initial strain amplitudes.  

This paper is organized as follows.
Section~\ref{sec:model} describes the simulation model.  
Sections~\ref{sec:oscillatory} and \ref{sec:stress_relaxation} present our
results for the two protocols.
Finally, we summarize our findings and conclude in Sec.~\ref{sec:sum}.

\section{Numerical modeling and Method}\label{sec:model}

We consider a two-dimensional equimolar binary mixture of frictionless particles with diameters 
$\DiameterS$ and $\DiameterL$, where the size ratio is
$\DiameterL/\DiameterS=1.4$, a value commonly used in studies of the jamming
transition~\cite{O'Hern2003pre,Silbert2005prl}. 
The interaction potential between particles $j$ and $k$ is given by
\begin{equation}
U(r_{jk}) = \frac{\epsilon}{\alpha}\left(1 - \frac{r_{jk}}{\Diameter_{jk}}\right)^{\alpha}
\Theta(\Diameter_{jk} - r_{jk}), 
\label{eq:potential}
\end{equation}
where $\Theta(x)$ is the Heaviside step function, 
$r_{jk}=|\bm{r}_j - \bm{r}_k|$, and 
$\Diameter_{jk}=(\Diameter_j+\Diameter_k)/2$, 
with $\Diameter_j$ and $\Diameter_k$ denoting the diameters of particles $j$ and
$k$, respectively~\cite{Durian1995prl}. 
In this study, we focus on the harmonic potential with $\alpha=2$.
The number of particles is $N = 1\,156$.
Throughout the simulations, we choose $\DiameterS$, 
$\epsilon$, and $\epsilon/\DiameterS^2$ as the units of length, energy, and stress, respectively.

We generate initial configurations via mechanical training using quasi-static
and cyclic volume changes, following the procedure of Kawasaki~\textit{et
al.}~\cite{Kawasaki2024prl}, in order to stabilize the jammed configurations
against shear deformation (see \ref{appendixA0} for details).
During the quasi-static deformation, we minimize the potential energy using the
FIRE algorithm~\cite{FIRE,Bitzek2006prl}. 
We regard the system as mechanically equilibrated when the magnitude of the
average force acting on a particle is less than $10^{-14}\epsilon/\DiameterS$. 
To eliminate residual stress in the initial configurations, we apply shear
stabilization~\cite{Kawasaki2024prl,Dagois-Bohy2012prl}, under Lees-Edwards
boundary conditions~\cite{Computer_Simulation_of_Liquids}. 
We first prepare a random configuration at $\phi = 0.8395$, and increase $\phi$ incrementally
by $\Delta\phi=10^{-4}$ while maintaining mechanical stability, up to the maximum 
packing fraction $\phi=0.9$.
Next, we decrease $\phi$ by $\Delta\phi =10^{-4}$ when 
$E = (1/N)\sum_{j>k} U(r_{jk})>10^{-8}$, 
and by $\Delta\phi =10^{-6}$ when $E<10^{-8}$.
We identify the packing fraction at which $E$ first satisfies $E < 10^{-16}$ as
the jamming transition point $\phiJ$. 

The dynamics of the jammed configuration under shear deformation are monitored
via molecular dynamics simulations. 
Throughout the paper, shear deformation is applied in the $x$-direction.
We assume overdamped particle dynamics with Stokes drag, so that the equation of
motion is given by
\begin{equation}
\zeta\dot{\bm{r}}_j(t) + \nabla_j \sum_{k<l} U(r_{kl}) = \bm{0},
\label{eq:EoM_shear}
\end{equation}
where $\zeta$ is the friction coefficient.
The unit of time is given by $t_0 = \zeta \DiameterS^2 / \epsilon$.
Under shear with a finite shear rate, the particle velocity
$\dot{\bm{r}}_j(t)$ in eqn~(\ref{eq:EoM_shear}) is replaced by
$\dot{\bm{r}}_j(t)-\dot{\gamma}(t) y_j(t)\bm{e}_x$, where $\bm{e}_x$ is the unit
vector in the $x$-direction. 
We numerically integrate eqn~(\ref{eq:EoM_shear}) using the Euler
method~\cite{Computer_Simulation_of_Liquids}. 
For the quasi-static process ($\omega\to0$), we use the numerical protocol of
Kawasaki \textit{et al.}~\cite{Kawasaki2024prl}: 
particle positions are relaxed to the mechanical equilibrium at every
incremental shear strain $\Delta\gamma$. 
For $\gamma<10^{-3}$, $\Delta\gamma$ is increased logarithmically from $10^{-9}$ to $10^{-3}$; 
for $\gamma>10^{-3}$, $\Delta\gamma$ is fixed at $10^{-3}$.
The shear stress is defined as~\cite{Computer_Simulation_of_Liquids} 
\begin{equation}
\sigma
=\frac{1}{L^2}\sum_{j>k}\frac{{x}_{jk}{y}_{jk}}{r_{jk}}\frac{\partial U}{\partial r_{jk}},
\label{eq:virial_stress}
\end{equation}
where $L$ is the system length, ${x}_{jk}=x_j-x_k$, and ${y}_{jk}= y_j-y_k$. 
All data presented in this paper are averages over at least 15 (typically 50) independent realizations.

\section{Oscillatory shear}\label{sec:oscillatory}

Consider the mechanical response of a jammed system under an oscillatory shear
strain defined as $\gamma(t)=\gamma\sin(\omega t)$, where $\gamma$ is the
strain amplitude and $\omega$ is the frequency. 
In the linear-response regime or $\gamma \to 0$, 
the complex modulus $G^*(\omega)=G'(\omega) + i G''(\omega)$ 
becomes independent of $\gamma$ 
and follows the scaling relation   
\begin{equation}
\Delta G'(\omega)=A\omega^{1/2}, ~~~(\omega > \omegas), 
    \label{eq:G_linear}
\end{equation}
where $\Delta G'(\omega) \equiv G'(\omega)-\GAQS$ and 
$\GAQS=G'(\omega\to0)$ is the static storage modulus. 
For the harmonic potential, $\GAQS$ scales as
$\dphi^{1/2}$,  
the onset frequency $\omegas$ scales as $\omegas\propto\dphi$, 
and the prefactor $A$ is \textit{independent} of $\dphi$~\cite{O'Hern2003pre,Tighe2011prl}. 
The loss modulus $G''(\omega)$ also follows the same scaling as eqn~(\ref{eq:G_linear}), 
except that the frequency window exhibiting the $\omega^{1/2}$ scaling extends
to lower frequencies than $G'(\omega)$~\cite{Hara2025natp}.  

As $\phi$ approaches $\phiJ$ from above, the linear-response regime narrows and 
softening sets in. 
In the quasi-static limit ($\omega\to 0$), 
we have found that the nonlinear static modulus $\GAQS(\gamma)$ 
follows a simple scaling law in the pre-yielding regime $\gamma < \gammay$~\cite{Kawasaki2024prl},  
\begin{equation}
\GAQS(\gamma)= \GAQS(\gamma\to0) \times \mathcal{G}\left(\frac{\gamma}{\gammas}\right),
    \label{eq:G_AQS}
\end{equation}
where $\mathcal{G}(x) = 1$ for $x\ll1$ and  $\propto x^{-1/2}$ for $x\gg1$.
The onset strain $\gammas \propto \dphi$ is independent of the exponent $\alpha$
in eqn~(\ref{eq:potential}). 
Thus, one finds 
\begin{equation}
\GAQS(\gamma) \sim \dphi \gamma^{-1/2}, ~~~(\gamma > \gammas) 
    \label{eq:G_nonlinear}
\end{equation}
for the harmonic potential~\cite{Kawasaki2024prl}. 
Note that, in both cases of $(\omega \neq 0, \gamma\to 0)$ and $(\gamma\neq 0, \omega\to 0)$, 
the moduli satisfy $G^{\ast}(\omega)=\GAQS(\gamma)=0$ below $\phiJ$. 

\begin{figure}[tb]
  \centering\includegraphics[width=0.95\columnwidth]{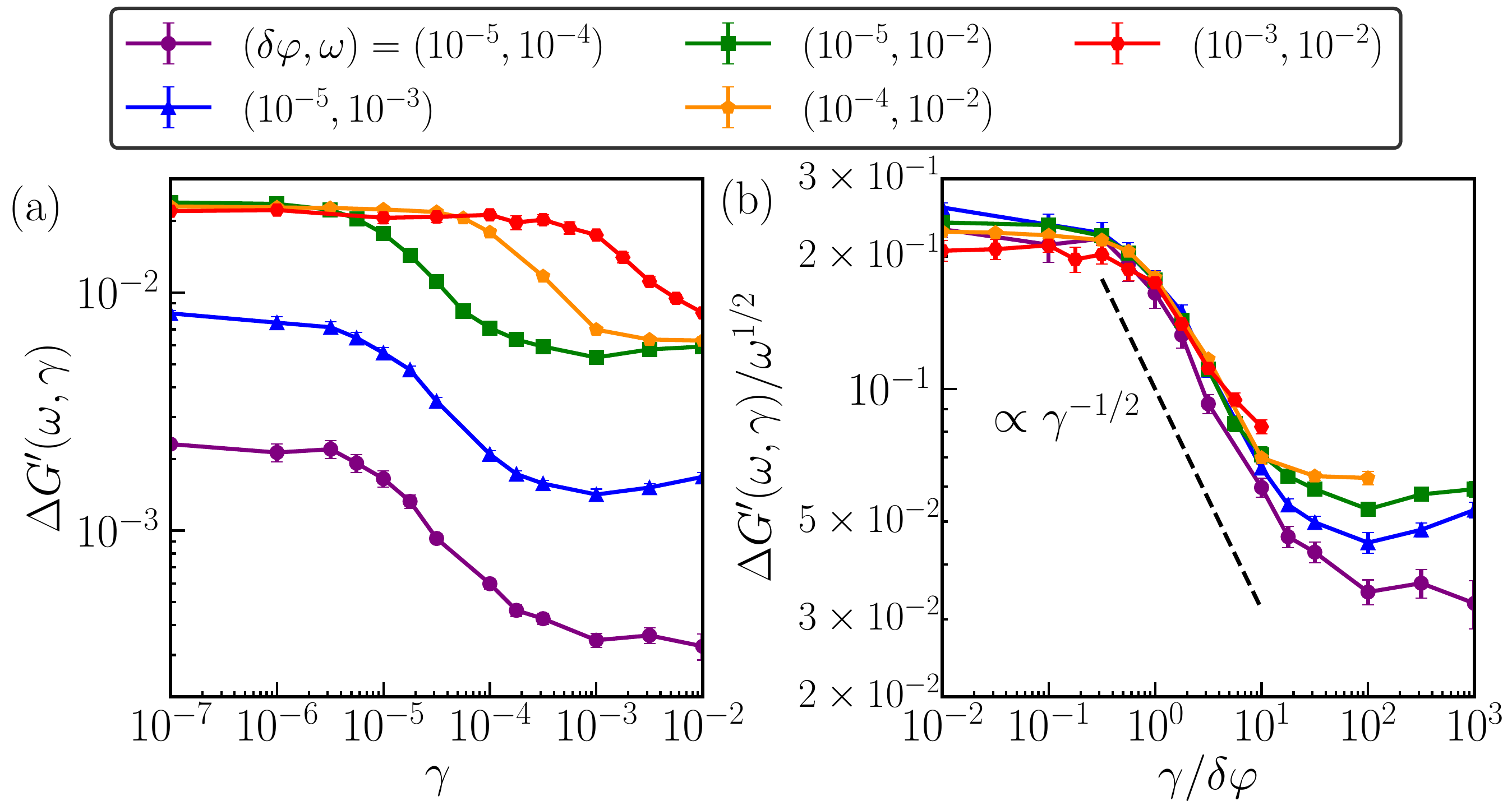}
  \caption{(a) $\gamma$-dependence of $\Delta G'(\gamma,\omega)$ for several
 sets of  $\dphi$ and $\omega$. 
(b) Scaled $\Delta G'(\omega)/\omega^{1/2}$ plotted as a function of $\gamma/\dphi$. 
The error bars in all panels represent the standard error of the mean.}
  \label{Fig1_G-vs-omega}
\end{figure}
In this section, we investigate the case in which both $\omega$ and $\gamma$ are finite 
and examine how the modulus $G^{\ast}(\omega, \gamma)$ depends on these parameters.
We systematically explore a broad range of $\omega$
within the $\omega^{1/2}$-scaling regime and consider systems both above and
below $\phiJ$. 

The storage and loss moduli are evaluated as~\cite{Otsuki2014} 
\begin{equation}
\displaystyle{G'(\omega, \gamma) = \frac{\omega}{\pi \gamma}
 \int_0^{2\pi/\omega}\dd t \sigma(t) \sin(\omega t),} 
    \label{eq:G_def}
\end{equation}
where $\sigma(t)$ is the stress in the stationary state. 
The loss modulus $G''(\omega, \gamma)$ is obtained by replacing
$\sin(\omega t)$ in eqn (\ref{eq:G_def}) with $\cos(\omega t)$. 
Starting from a jammed packing generated using the protocol described in the
previous section, we subject the system to multiple oscillatory shear cycles until 
the particle trajectories become periodic,
ensuring that the system has reached the stationary state. 

We first consider the case of $\phi > \phiJ$. 
We begin by confirming that $G^{\prime}(\omega, \gamma\to 0)=G^{\prime}(\omega)$ 
in the linear-response regime follows the $\omega^{1/2}$ scaling given by
eqn~(\ref{eq:G_linear})~\cite{Baumgarten2017sm}, and that 
$G'(\omega\to 0, \gamma) =\GAQS(\gamma)$ in the quasi-static limit 
follows the $\gamma^{1/2}$ scaling in  eqn~(\ref{eq:G_nonlinear})~\cite{Kawasaki2024prl}
(see Fig.~\ref{figA1_Gomega_GAQS} in \ref{appendixA}).
Next, we examine the dependence of the modulus on $\gamma$ 
as $\omega$ increases from zero to a finite value larger than $\omegas$. 
As discussed below eqn~(\ref{eq:G_linear}), the prefactor $A$ is independent of $\dphi$ 
for the harmonic potential. 
The question is how $A$ is modified with increasing $\gamma$ in the 
softening regime. 
Figure~\ref{Fig1_G-vs-omega}(a) shows $\Delta G'(\omega,\gamma)$ as a
function of $\gamma$ for various $\dphi$ and $\omega$. 
The results show that $\Delta G'(\omega, \gamma)$ decreases with increasing 
$\gamma$ at fixed $\omega$, a clear signature of softening,  
and increases with increasing $\omega$ at fixed $\gamma$.
The softening saturates around $\gamma\approx10^{-4}$.
A similar trend is observed for $\GAQS(\gamma)$ in the quasi-static
limit (see Fig.~\ref{figA1_Gomega_GAQS}(b)). 
Kawasaki \textit{et al.}~\cite{Kawasaki2024prl} have shown that this behavior
arises from strain hardening when the initial jammed configurations are prepared
by mechanical training.
In Fig.~\ref{Fig1_G-vs-omega}(b), we plot the scaled quantity
$\Delta G'(\omega,\gamma)= G'(\omega,\gamma)- \GAQS(\gamma)$ 
normalized by $\omega^{1/2}$ as a function of $\gamma/\dphi$. 
The collapse of the data indicates that the prefactor $A$ in eqn~(\ref{eq:G_linear}) remains 
independent of $\omega$ but becomes a nonlinear function of $\gamma$. 
The shear-softening exponent is found to be approximately $-1/2$.
A qualitatively similar result for $G''(\omega,\gamma)$ is shown in
Fig.~\ref{FigA2_G2-vs-gamma} in \ref{appendixA}, although the softening exponent is slightly smaller.
We attribute this to a small deviation from the $\omega^{1/2}$ scaling 
of $G''(\omega)$ at high frequencies already seen in the linear-response regime as shown in 
Fig.~\ref{figA1_Gomega_GAQS}(a)~\cite{Baumgarten2017sm}.

\begin{figure}[tb]
    \centering
\includegraphics[width=0.98\columnwidth]{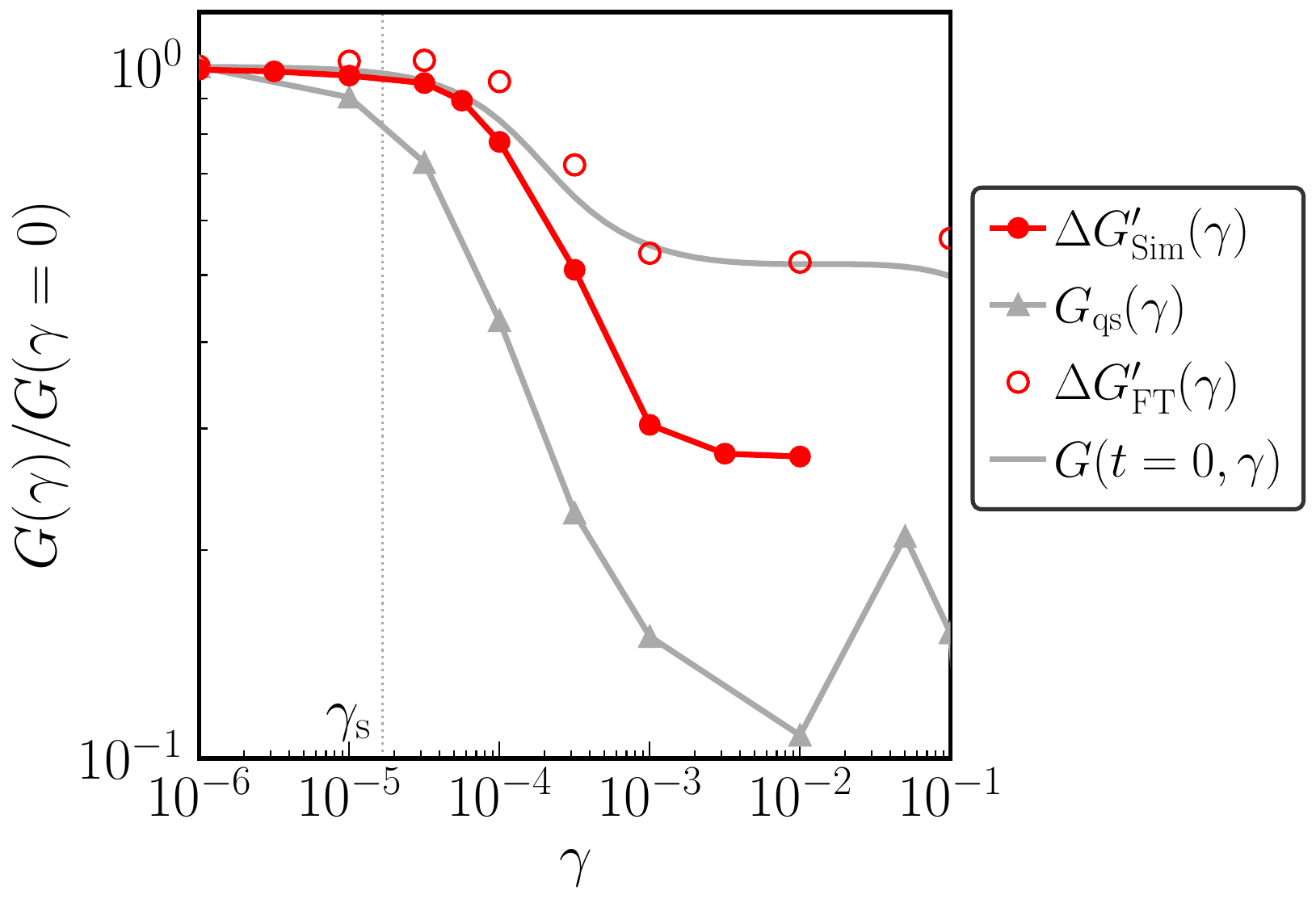}
 \caption{
$\gamma$-dependence of the moduli normalized by their values at $\gamma=0$ for
 $\dphi=10^{-4}$.  
$\Delta G'_{\mathrm{Sim}}(\omega,\gamma)$ (red filled circles) at
 $\omega=10^{-2}$ taken from Fig.~\ref{Fig1_G-vs-omega}.
$\GAQS(\gamma)$ (gray filled triangles) from Fig.~\ref{figA1_Gomega_GAQS}(b). 
$\Delta G'_{\mathrm{FT}}(\omega, \gamma)$ (open circles) evaluated by 
the Fourier transform of $G(t)$ shown in Fig.~\ref{Fig5_stressrelaxation}(a)
 (see the next section). 
$G(t=0)$ (gray solid line) is taken from Fig.~\ref{Fig5_stressrelaxation}(a). 
The thin vertical line indicates $\gammas$ for $\GAQS(\gamma)$. 
}
    \label{Fig2_Gqs-GFT-G0}
\end{figure}
Figure~\ref{Fig1_G-vs-omega}(b) demonstrates 
that $\Delta G^{\prime}(\omega, \gamma)$ follows both the $\omega^{1/2}$ scaling 
of eqn~(\ref{eq:G_linear}) and the $\gamma^{1/2}$ scaling of eqn~(\ref{eq:G_AQS}) simultaneously,
implying the following scaling form:
\begin{equation}
\Delta G'(\omega, \gamma)=A\omega^{1/2} 
\times \mathcal{G}_{\omega}\left(\frac{\gamma}{\gammas}\right).
    \label{eq:G_omega_gamma0}
\end{equation}
We find that the scaling function $\mathcal{G}_{\omega}(x)$ is similar to,
but not identical to $\mathcal{G}(x)$ defined in eqn~(\ref{eq:G_AQS}). 
In other words, $\Delta G^{\prime}(\omega, \gamma)$ 
is not simply a superposition of eqns~(\ref{eq:G_linear}) and (\ref{eq:G_AQS}). 
In Fig.~\ref{Fig2_Gqs-GFT-G0}, we plot the two moduli normalized by their values at
$\gamma=0$; 
$\mathcal{G}_{\omega} =\Delta G'(\omega, \gamma)/A\omega^{1/2}$ 
(red filled circles) and 
$\mathcal{G}= \GAQS(\gamma)/\GAQS$ (gray filled triangles). 
The packing fraction is $\dphi =10^{-4}$. 
The figure shows that the two moduli follow similar scaling forms   
and deviate from them at a common strain value around $\gamma \gtrsim 10^{-4}$. 
However, the onset strains $\gammas$ differ.
For $\GAQS(\gamma)$, we find $\gammas \approx 1.7\times 10^{-5}$, which is substantially smaller 
than $\gammas \approx 6 \times 10^{-5}$ for $\Delta G^{\prime}(\omega, \gamma)$.
This can be understood as follows.
The fact that $\gammas$ is proportional to $\dphi$ for both quantities (as shown 
in Figs.~\ref{Fig1_G-vs-omega}(b) and \ref{figA1_Gomega_GAQS}(b)) suggests a
common physical mechanism underlying the softening. 
In the quasi-static limit, the softening occurs when the particle's non-affine displacement
under strain becomes sufficiently large to release two particles' overlaps~\cite{Kawasaki2024prl}.   
The same mechanism should also operate at finite $\omega$.
However, under oscillatory shear with finite $\omega$, a substantial fraction of
particle motion is affine-like.
Configurations generated predominantly by affine displacements are less efficiently packed
compared to those arising from non-affine displacements, which dominates in
the quasi-static deformation. 
Therefore, the breaking of particle's bonds and release of the particle's overlaps occur at a smaller
strain in the quasi-static case than in oscillatory shear. 
In the next section, we discuss the affine contribution to the
softening in detail (see also other data in Fig~\ref{Fig2_Gqs-GFT-G0}).

\begin{figure}[tb]
    \centering    \includegraphics[width=0.8\columnwidth]{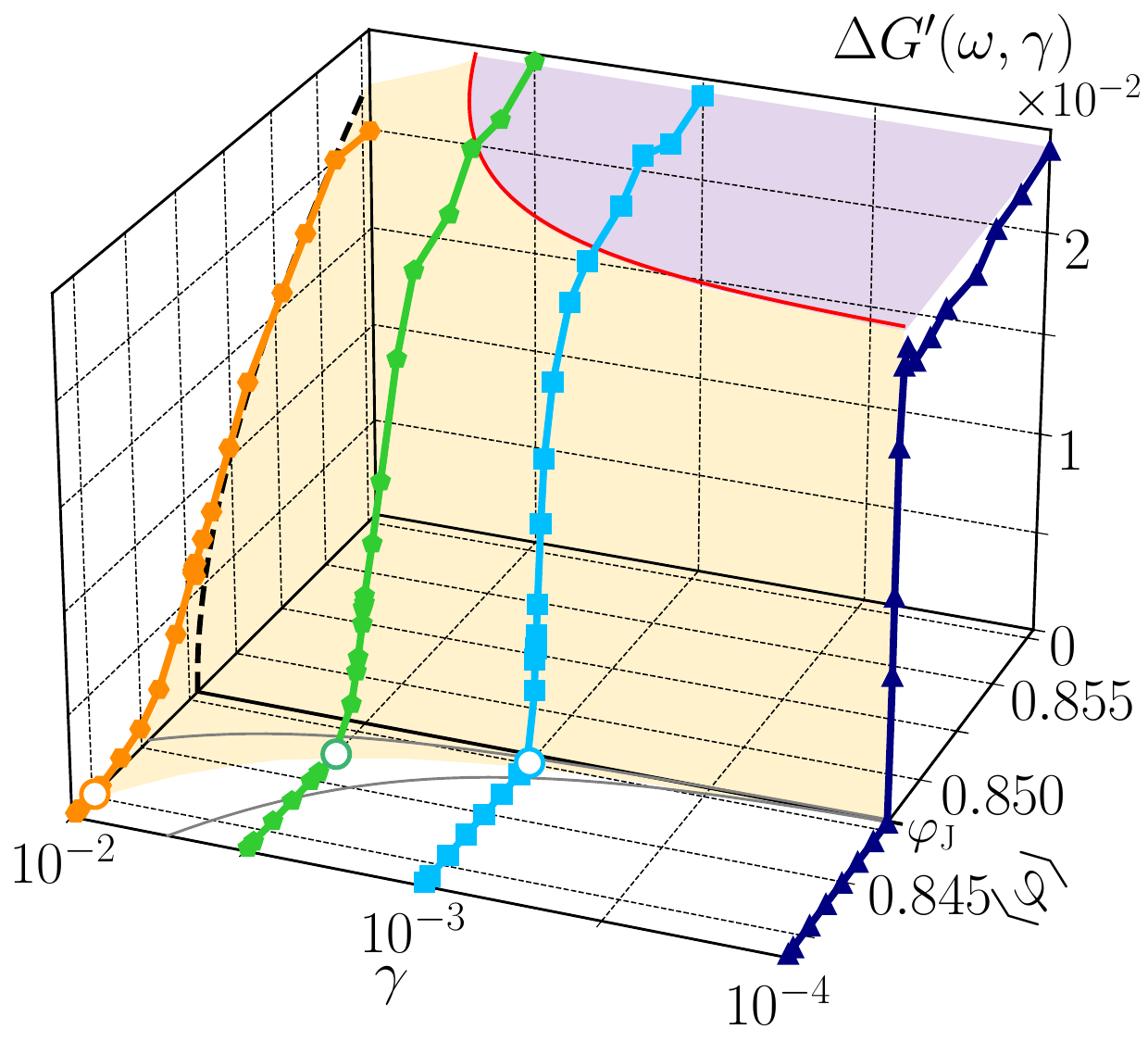}
    \caption{
    $\Delta G'(\omega, \gamma)$ as a function of $\ev{\phi}$ and $\gamma$ 
at a fixed frequency of $\omega=10^{-2}$. 
    The thin black line on the $\Delta G'(\omega,\gamma)=0$-plane represents the
 jamming transition point $\phiJ\approx0.8455$, while the thin red line
 indicates the onset strain of softening, $\gammas$. 
    The symbols denote the modulus at $\gamma=10^{-4}$ (navy triangles), $10^{-3}$ (light blue squares), $10^{-2.5}$ (green pentagons), and $10^{-2}$ (orange hexagons). 
    The dashed line at $\gamma=10^{-2}$ represents a fit to $\dphi^{1/2}$. 
    The open circles indicate the packing fraction $\phi_0$ at which $\Delta
 G'(\omega,\gamma)$ vanishes.  
    The thin gray lines on the $\Delta G'(\omega,\gamma)=0$ plane connect
 $\phiJ$ and the data for the point-to-loop reversible transition line closest
 to $\phiJ$ reported in Schreck \textit{et al.}~\cite{Schreck2013pre} (one
 between $\phiJ$ and $\phi_0$) and Nagasawa \textit{et
 al.}~\cite{Nagasawa2019sm} (one below $\phi_0$).  
}
    \label{Fig3_3d-G}
\end{figure}
If the scaling relation in eqn~(\ref{eq:G_omega_gamma0}) remains valid down to $\phiJ$, 
$\Delta G'(\omega, \gamma)$ should behave as 
\begin{equation}
\Delta G'(\omega, \gamma) \sim 
\left(\omega  \dphi/\gamma\right)^{1/2}, ~~~
(\phi \to \phiJ+)
    \label{eq:G_omega_gamma0-2}
\end{equation}
for the harmonic potential.
In other words, the modulus is expected to vanish at $\phi = \phiJ$. 
Indeed, this is the case when $\omega\neq 0$ with $\gamma \to 0$ or 
$\gamma\neq 0$ with $\omega \to 0$.
However, as shown below, this no longer holds when both $\omega$ and  $\gamma$ are finite. 
The scaling relation in eqn~(\ref{eq:G_omega_gamma0}) breaks down 
in the vicinity of $\phiJ$ and 
the modulus remains finite even below $\phiJ$.  
Furthermore, the stress is no longer linear in strain at $\phi \leq \phiJ$. 
This can be understood by noting that, for athermal, non-Brownian, and overdamped particles, 
no collisions occur during an oscillatory shear cycle if $\gamma$ is sufficiently small.
Only when $\gamma$ exceeds a threshold value 
do particles begin to collide with
their neighbors, producing a finite stress.

Figure~\ref{Fig3_3d-G} shows $\Delta G'(\omega,\gamma)$ as a function 
$\phi$ 
for several values of $\gamma$ at a fixed frequency $\omega=10^{-2}$. 
For the $\phi$-axis, we use $\ev{\phi}\equiv \ev{\phiJ}+\dphi$, where  
$\ev{\phiJ}$ is the sample-averaged jamming transition point, 
since $\dphi$ is the control parameter throughout this study.
The range of $\ev{\phi}$ shown in the figure corresponds to
$-6\times10^{-3} \leq \dphi \leq 10^{-2}$.
The figure demonstrates that the scaling relation in eqn~(\ref{eq:G_linear}) holds in
the limit $\gamma\to0$: at $\gamma=10^{-4}$, $\Delta G'(\omega, \gamma)$ is independent of 
$\phi$ above $\phiJ$ and drops discontinuously to zero below $\phiJ$.
As $\gamma$ increases, this discontinuous drop becomes increasingly rounded.
As predicted by eqn~(\ref{eq:G_omega_gamma0-2}),  
$\Delta G'(\omega,\gamma) \propto \dphi^{1/2}$ 
in the softening regime ($\gamma > \gammas$ ) at fixed $\omega$, as 
indicated by the black dashed line in Fig~\ref{Fig3_3d-G} at $\gamma=10^{-2}$. 
However, as $\phi$ approaches $\phiJ$ for fixed $\gamma$, 
$\Delta G'(\omega,\gamma)$ deviates from eqn~(\ref{eq:G_omega_gamma0-2}), 
crosses smoothly through $\phiJ$, and vanishes at a packing fraction below $\phiJ$.
The packing fraction $\phi_0$, at which $\Delta G'(\omega,\gamma)$ reaches zero, 
decreases with increasing $\gamma$.
We find that the modulus obeys a simple algebraic form:
\begin{equation}
    \Delta G'(\omega, \gamma) \sim(\phi-\phi_0)^{\beta}.
\end{equation}
Our preliminary simulations yield a positive exponent $\beta \approx 1.2$ for $G'(\omega, \gamma)$ at $\gamma=10^{-2}$. 
$\beta$ appears to vary with $\gamma$ and $\omega$, while
it is independent of $\phi_0$. 
For the loss modulus $G''(\omega, \gamma)$, the exponent is larger  
(Fig.~10 in Appendix B). 
A detailed analysis is needed, but it is challenging because the
relaxation toward stationary states becomes prohibitively slow and
is expected to depend sensitively on the system size near $\phi_0$. 
We leave this important problem for future work.

The emergence of stress at $\phi \geq \phi_0$
is caused by transient contact networks formed by colliding particles under oscillatory shear. 
The emergence of such networks with increasing $\gamma$ below $\phiJ$
has been studied in the context of 
the reversible-irreversible (or absorbing) transition 
of particle
trajectories~\cite{Milz2013pre,Schreck2013pre,Nagasawa2019sm,Matsuyama2021epje,Das2010pnas}. 
These studies show that a system subjected to oscillatory shear 
exhibits distinct dynamical phases. 
For small $\gamma$, particles move affinely and return to their initial positions after each cycle. 
This small $\gamma$-regime is called \textit{point-reversible} phase, as the particles 
remain immobile in the co-moving frame.
As $\gamma$ increases, particles begin to interact with their neighbors and
follow non-affine trajectories, yet still return to their original positions
after each cycle. 
This regime is referred to as the \textit{loop-reversible} phase. 
At even larger $\gamma$, the particles diffuse irreversibly and do not
return to their original configurations, defining the \textit{irreversible}
phase~\cite{Milz2013pre,Schreck2013pre,Nagasawa2019sm,Matsuyama2021epje,Das2010pnas}. 
Since the transition point from the point-reversible to loop-reversible phases, $\phi_{\mathrm{PL}}$, 
is the density at which the stress-bearing contact networks percolate,  it is expected that $\phi_{\mathrm{PL}}$ is the point at which the elastic response becomes finite, \textit{i.e.}, $\phi_{\mathrm{PL}}=\phi_0$.
The two thin gray lines on the $\Delta G'(\omega,\gamma)=0$-plane 
in Fig.~\ref{Fig3_3d-G} connect $\phiJ$ with the available data for
the point-to-loop reversible transition 
reported in Schreck \textit{et al.}~\cite{Schreck2013pre} and Nagasawa \textit{et
al.}~\cite{Nagasawa2019sm}. 
All values of $\phi_0$ (open circles) lie between these two lines. 
The comparison remains semi-quantitative due to the limited resolution of earlier
studies near
$\phiJ$~\cite{Milz2013pre,Schreck2013pre,Nagasawa2019sm,Matsuyama2021epje,Das2010pnas}.   
A more precise determination of $\phi_{\mathrm{PL}}$ ($=\phi_0$), together with
a detailed characterization of the critical behavior and its frequency dependence, is needed.

\section{Transient stress relaxation}\label{sec:stress_relaxation}
In this section, we investigate the transient stress relaxation of jammed
packings, which represents the real-time analogue of 
the oscillatory-shear measurement and thus complements the results discussed in the previous section. 
A step shear strain $\gamma$ is applied to the system at time $t=0$, and the
resulting shear stress $\sigma(t)$ and other observables are
monitored as functions of time $t\;(>0)$.
For sufficiently small $\gamma$, where the linear-response approximation holds, 
the constitutive equation for the shear stress can be written as
\begin{equation}
    \sigma(t) = \int_{-\infty}^t\dd t'~ G(t - t') \dot{\gamma}(t'),
    \label{eq:linear_viscoelastic_theory}
\end{equation}
where $\dot{\gamma}(t)$ is the shear rate and $G(t)$ is the relaxation modulus. 
The Fourier transform of $G(t)$ is related to the complex modulus $G^*(\omega)$
by~\cite{larson1999structure}: 
\begin{equation}
    G^*(\omega)= \GAQS + i\omega\int^{\infty}_{0}\dd t~e^{-i\omega t}\Delta G(t),
    \label{eq:FT_Gt}
\end{equation}
where $\Delta G(t)= G(t) -\GAQS$.
For a step strain, where $\dot{\gamma}(t)=\gamma\delta(t)$, 
eqn~(\ref{eq:linear_viscoelastic_theory}) reduces to $\sigma(t)=G(t)\gamma$.
These equations imply that the $\omega^{1/2}$ scaling, 
$G^*(\omega)\sim A\omega^{1/2}$ in eqn~(\ref{eq:G_linear}) corresponds
to an algebraic decay of the stress:
\begin{equation}
    G(t)=\sigma(t)/\gamma\sim t^{-1/2}.
\label{eq:t-decay}
\end{equation}
The stress relaxation is expected to persist up to
$t_{\mathrm{s}}= \omegas^{-1}$, the inverse of the onset frequency
of the $\omega^{1/2}$-scaling regime.  
Such scale-free relaxation has been observed 
near the jamming transition $\phiJ$~\cite{Hatano,Saitoh2020prl}. 

\begin{figure}[tb]
  \centering
  \includegraphics[width=0.95\columnwidth]{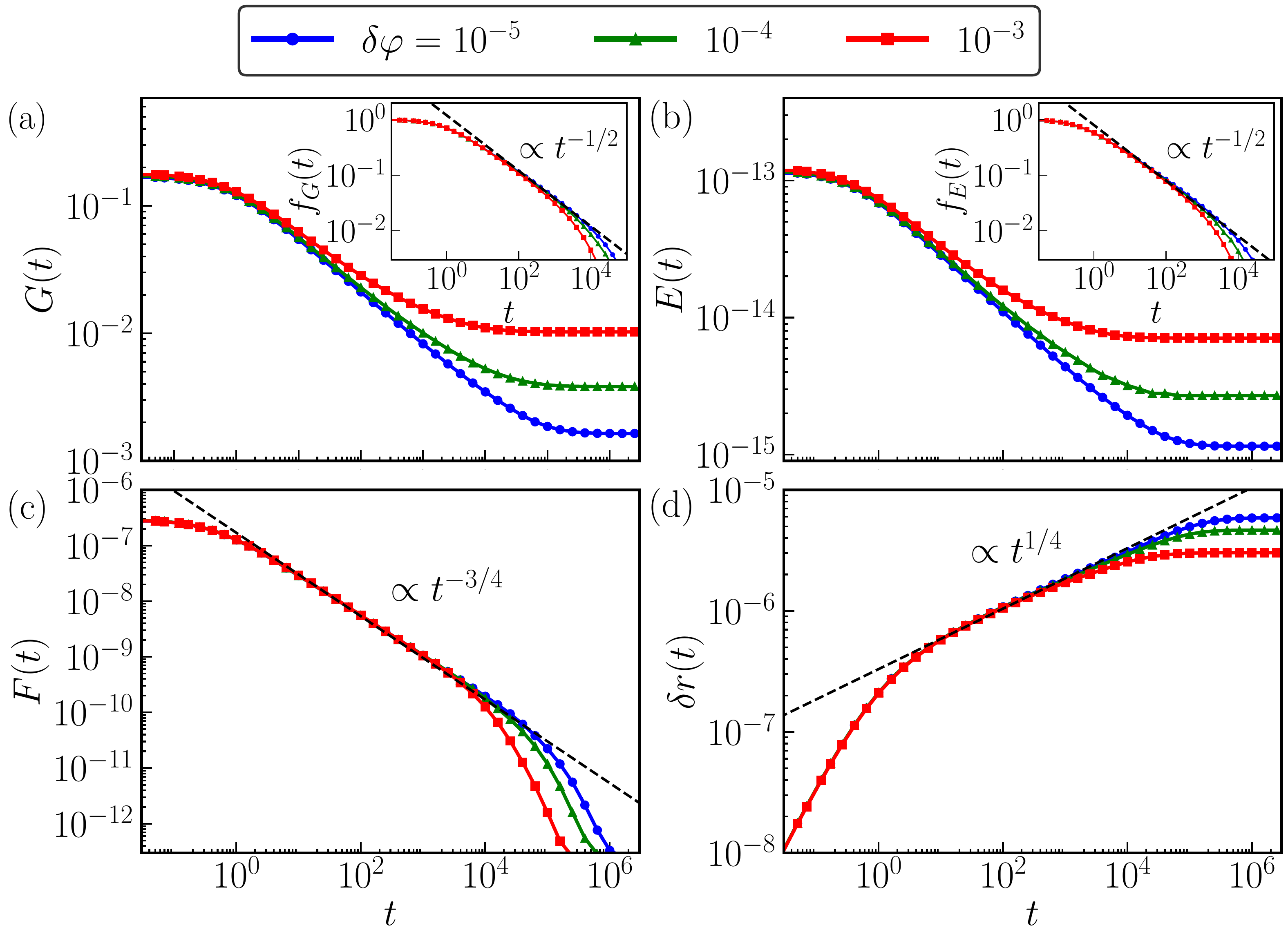}
  \caption{
Transient relaxation in the linear-response regime at a fixed strain $\gamma=10^{-6}$ 
for $\dphi=10^{-5}$ (blue), $10^{-4}$ (green), and $10^{-3}$ (red). 
(a) Relaxation modulus $G(t)$. 
The inset shows the normalized function $f_G(t)$.
(b) Potential energy $E(t)$. The inset shows $f_E(t)$.
(c) Average force $F(t)$.
(d) Non-affine particle displacement $\Dr(t)$.
The dashed lines indicate the algebraic scaling. 
}
  \label{Fig4_linear_stressrelaxation}
\end{figure}
The main question addressed in this section is how the algebraic decay in 
eqn~(\ref{eq:t-decay}) changes with increasing strain amplitude $\gamma$. 
Recall that the algebraic behaviors of $G^*(\omega)\sim\omega^{1/2}$ and
$G(t)\sim t^{-1/2}$ observed in the linear-response regime can be explained,
within the harmonic approximation, as the superposition of the exponential
relaxation modes of a jammed packing within a single harmonic basin of the
potential energy landscape~\cite{Tighe2011prl}.   
With increasing $\gamma$ beyond the onset strain $\gammas$, 
nonlinear response, or softening sets in.
In this regime, the harmonic approximation is expected to break down, 
and the relaxation of the initial configuration toward the terminal state
likely involves traversing saddles in the rugged energy landscape 
rather than a superposition of relaxations in a single basin. 
This naturally lead to deviations from the $G(t)\sim t^{-1/2}$ behavior. 
If $\gamma$ increases further,  the configuration becomes so strongly distorted from the
initial jammed packing that the resulting relaxation dynamics are expected to resemble
those of completely random configurations.  
This is analogous to the steepest-descent dynamics of particle systems
quenched from high temperature $T=\infty$ to $T=0$, which have been recently 
studied by simulations~\cite{Chacko2019prl,Ikeda2020prl,nishikawajstatphys2021,Nishikawa2022prx}.   
These studies report that observables such as the energy $E(t)$ decay algebraically with
a nontrivial exponent $\beta$ ranging between 0.7 and 1.3 depending on the
spatial dimension and on the temperature at which initial configurations are
prepared~\cite{Chacko2019prl,Nishikawa2022prx}.   

Below, we evaluate $G(t)$ and $E(t)$, together with several other 
observables, over a broad range of $\gamma$ and show that 
$G(t)$ and $E(t)$ decay as $t^{-1/2}$ at $\gamma <\gammas$
and  $E(t)$ decays as $t^{-\beta}$ at $\gamma \gg \gammay$ (where $\gammay$ is the yielding threshold) with the same exponent $\beta$ as reported 
for the steepest-descent dynamics from $T=\infty$~\cite{Chacko2019prl,Nishikawa2022prx}.  
Surprisingly, we find that $G(t) \sim t^{-1/2}$ 
holds 
even in the softening regime, 
at $\gammas < \gamma  <\gammay$.  
We show that this apparently robust exponent $1/2$ in the softening region
cannot be fully accounted for by the harmonic approximation.
The dependence of the transient stress relaxation on $\gamma$ has been investigated
by pioneering work by Boschan \textit{et al.}~\cite{Boschan2016sm,Boschan2}. 
Here, we extend their analysis and explore a much broader
range of $\gamma$, thereby providing continuous connection between 
the linear-response regime and large-$\gamma$ limit.

In our simulation, all particle positions are instantaneously sheared at
$t=0$, such that  $(x_j,y_j) \to (x_j+\gamma y_j,y_j)$ for $j=1,\cdots, N$.
This sudden deformation introduces a force imbalance in the system.
We then monitor the relaxation of the particles toward a new mechanical
equilibrium for $t>0$, while maintaining the applied strain.
The particle dynamics follow the overdamped equation of motion given by eqn~(\ref{eq:EoM_shear}).
We compute several relevant physical observables: 
the shear stress or the relaxation modulus, $G(t)=\sigma(t)/\gamma$, 
the potential energy change (with respect to the unsheared system)
$E(t)\equiv \frac{1}{N}\sum_{j>k}U(r_{jk})-E_{\gamma=0}$, 
where $E_{\gamma=0}$ is the energy before applying the step strain, 
the average magnitude of the force, $F(t)\equiv \frac{1}{N}\sum_{j=1}^N|\bm{F}_j|$, 
and the non-affine particle displacement, 
$\Dr(t) \equiv \frac{1}{N}\sum_{j=1}^N|\bm{r}_j(t)-\bm{r}_j(0)|$.

\begin{figure}[tb]
  \centering
  \includegraphics[width=0.76\columnwidth]{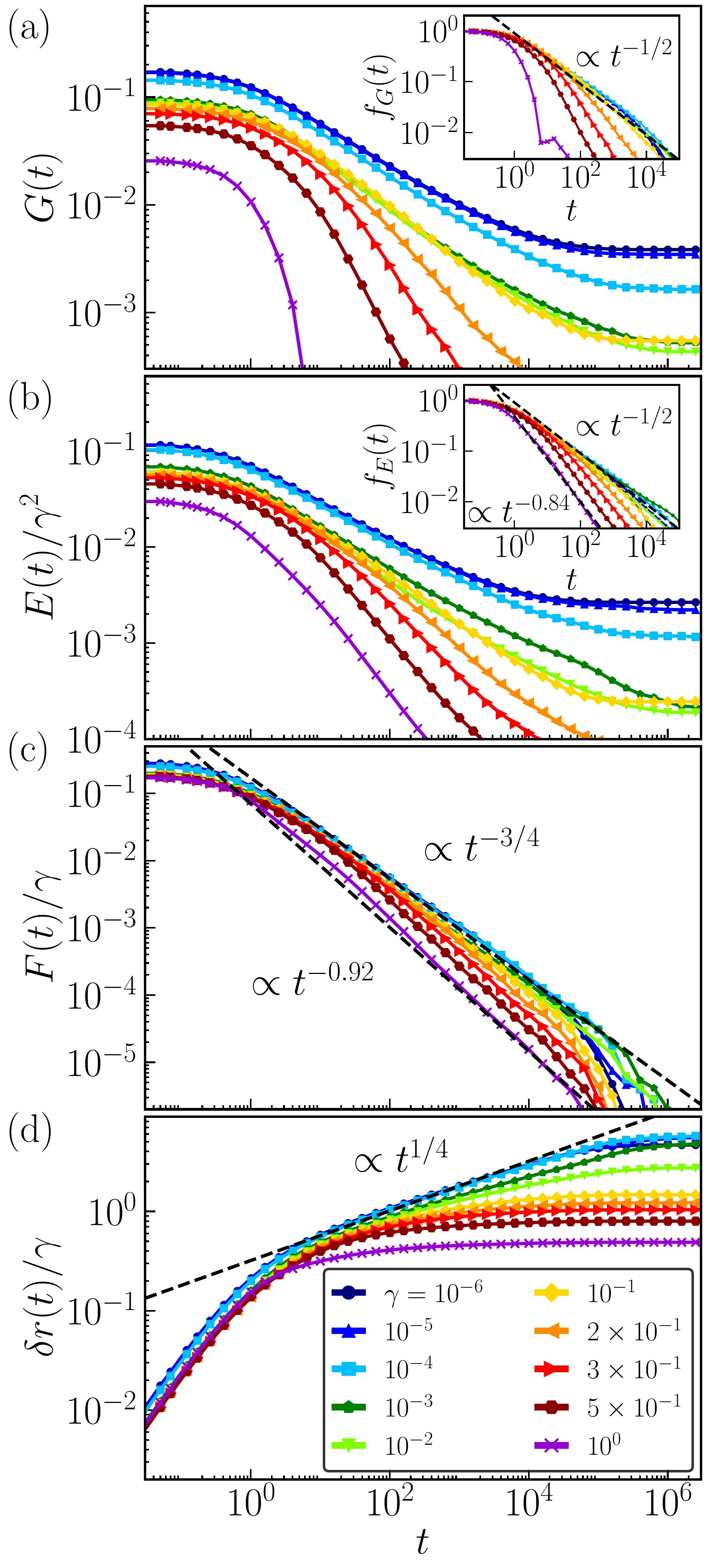}
  \caption{
The transient relaxation for various $\gamma$ (see the legend)
at $\dphi=10^{-4}$. 
(a) $G(t)=\sigma/\gamma$, 
(b) $E(t)/\gamma^2$,
(c) $F(t)/\gamma$, 
and (d) $\Dr(t)/\gamma$. 
The insets in (a) and (b) show the time evolution of the normalized
functions $f_{\calO}(t)$. 
The dashed lines represent the algebraic scaling.}
  \label{Fig5_stressrelaxation}
\end{figure}
We first confirm the transient relaxation in the linear-response regime,
consistent with previous studies~\cite{Hatano,Boschan2,Saitoh2020prl}. 
Figure~\ref{Fig4_linear_stressrelaxation} shows the observables for three
values of $\dphi$ at a fixed value of $\gamma=10^{-6}$, which is
sufficiently small to ensure linear response.
Figure~\ref{Fig4_linear_stressrelaxation}(a) and (b) 
show that  both $G(t)$ and $E(t)$ decay as
$t^{-1/2}$ at long times before eventually saturating at large $t$~\cite{Boschan2,Hatano,Saitoh2020prl}. 
The insets present the normalized relaxation function, defined as 
\begin{equation}
f_{\calO}(t) \equiv 
\frac{\calO(t)-\calO(\infty)}{\calO(0)-\calO(\infty)}, 
\label{eq:normalized-f}
\end{equation}
where $\calO= G, ~E$.
In Fig.~\ref{Fig4_linear_stressrelaxation}(a),
$f_{G}(t)$ demonstrates that the prefactor of the $t^{-1/2}$ decay is independent of $\dphi$, 
in agreement with the $\omega^{1/2}$ scaling predicted by eqn~(\ref{eq:G_linear}).
We also confirm that $G(t\to\infty)$ coincides with $\GAQS$. 
The relaxation of $E(t)$ in Fig.~\ref{Fig4_linear_stressrelaxation}(b) is
essentially identical to that of $G(t)$, since $E(t) \sim G(t)\gamma^2$ in the linear-response regime. 
These algebraic behaviors can be rationalized within the harmonic approximation, and
the detailed derivations of these relations are summarized in \ref{appendixB}.
Figure~\ref{Fig4_linear_stressrelaxation}(c) and (d) show 
that the average force relaxes as $F(t) \sim t^{-3/4}$, while  the non-affine
particle displacement increases as $\Dr(t)\sim t^{1/4}$.
These exponents can be understood by noting that the interparticle force 
is balanced with the Stokes drag, $F(t) \sim \zeta \dot{r}(t)$.
Substituting this to the energy dissipation rate, $d E/d t = -\zeta \dot{r}^2(t)$, 
one obtains $F(t) \sim t^{-3/4}$.
Similarly, integrating the velocity over time gives $\Dr(t) \sim t^{1/4}$. 

\begin{figure}[tb]
  \centering
  \includegraphics[width=0.95\columnwidth]{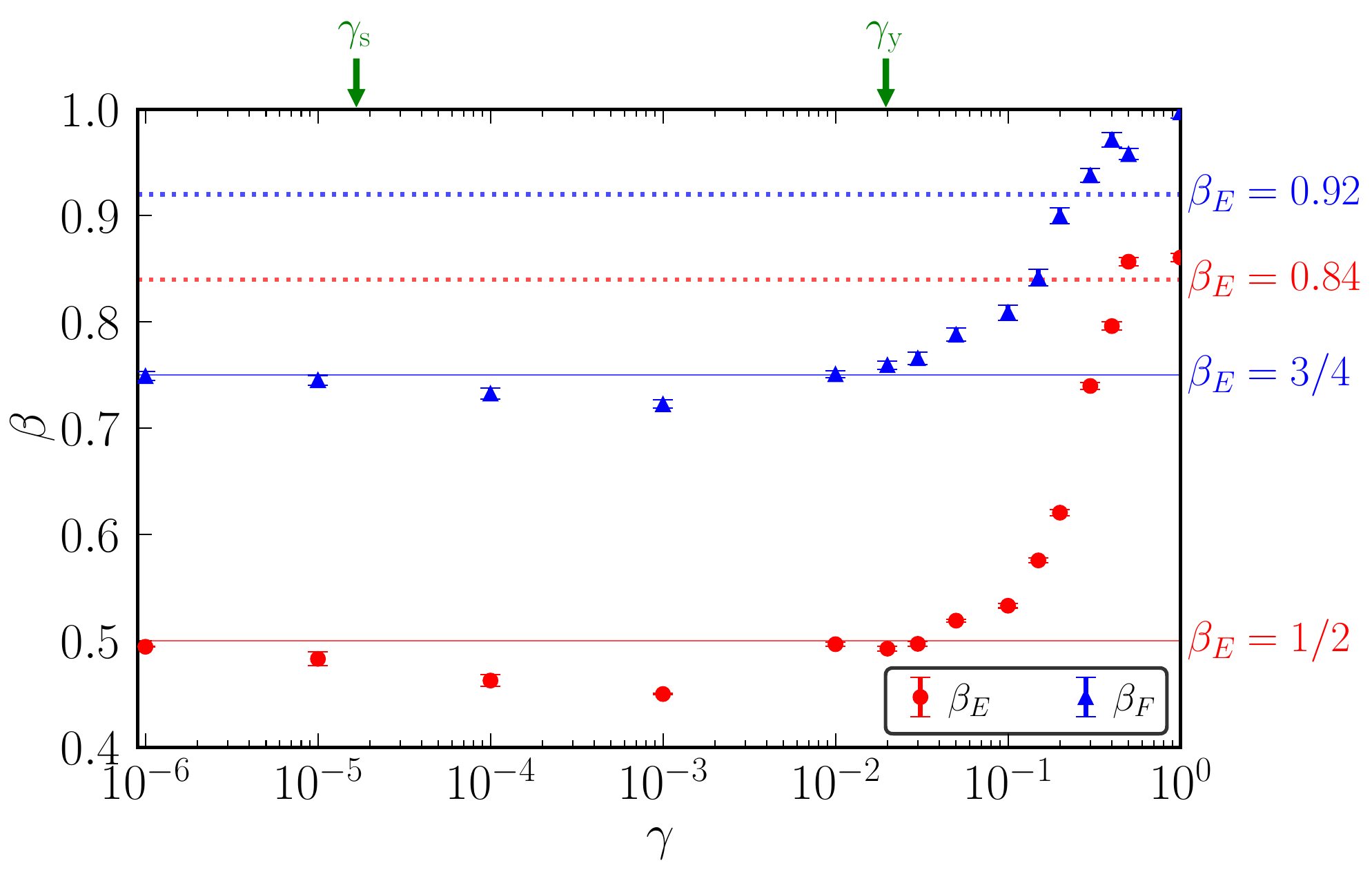}
  \caption{$\gamma$-dependence of the exponents $\beta_E$ and $\beta_F$ 
for the algebraic relaxation of $E(t)$ (red circles) and $F(t)$ (blue triangles)
 with error bars representing the standard error, respectively. 
The horizontal solid lines represent the linear-response values,
 $\beta_E=1/2$ and $\beta_F=3/4$. 
The horizontal dotted lines denote the values reported for the steepest-descent
 dynamics in systems quenched from $T=\infty$ to $T=0$~\cite{Chacko2019prl,Nishikawa2022prx}: 
 $\beta_F=0.92$ and $\beta_E= 2\beta_F-1=0.84$. 
The two arrows at the top of the panel indicate the onset strain $\gammas$ for
 $\GAQS(\gamma)$ and the yielding threshold $\gammay$.
 } 
  \label{Fig6_beta-vs-gamma}
\end{figure}
Now, let us systematically investigate how the transient relaxations evolve with increasing $\gammaz$. 
Figure~\ref{Fig5_stressrelaxation} shows the time evolution of the observables 
for a broad range of $\gammaz$ from $\gammaz=10^{-6}$, well within the
linear-response regime, up to $\gammaz=1$, which extends far beyond the yielding
threshold $\gammay\approx 2\times 10^{-2}$.   
The packing fraction is fixed at $\dphi=10^{-4}$.  
The quantities $F(t)$ and $\Dr(t)$ are scaled by $\gammaz$, 
and $\Delta E(t)$ by $\gammaz^2$, such that all observables become independent of $\gammaz$ 
in the linear-response regime.
We confirm that $G(t\to\infty)$ in Fig.~\ref{Fig5_stressrelaxation}(a)
agrees with the quasi-static modulus  $\GAQS(\gammaz)$ in Fig.~\ref{figA1_Gomega_GAQS}(b). 
These  results reveal several nontrivial features.

First, $G(t)$ and $E(t)/\gamma^2$ at short times decrease with increasing $\gamma$. 
In other words, not only the quasi-static values $G(t\to\infty)=\GAQS(\gamma)$ and
$E(t\to\infty)/\gamma^2$, but also the instantaneous responses $G(t=0)$ and
$E(t=0)/\gamma^2$ 
exhibit shear softening. 
The short-time response to shear is dominated by the \textit{affine}
displacements of the constituent particles.  
Our results indicate that the deformation is affine, but that the response becomes progressively nonlinear.
The $\gamma$-dependence of $G(t=0)$ is plotted as the solid gray line in
Fig.~\ref{Fig2_Gqs-GFT-G0}. 
It shows that the softening of $G(t=0)$ is weaker than that of $\GAQS(\gamma)$
and $\Delta G'(\omega,\gamma)$. 
Interestingly, the onset strain for softening $\gammas$ for $G(t=0)$ is comparable to
that of $\Delta G'(\omega,\gamma)$. 
We also confirm that $\gammas$ for $G(t=0)$ scales as $\dphi$ (results not shown). 
This observation further supports the argument in the previous section that 
the affine
contribution to $\Delta G'(\omega,\gamma)$ shifts the 
onset strain to a larger value than that of $\GAQS(\gamma)$.

Second, and more interestingly, all observables exhibit an algebraic
time dependence over the entire range of $\gamma$, except for $G(t)$ in the
large-$\gamma$ regime beyond $\gammay$. 
We focus in particular on $E(t)\sim t^{-\beta_E}$ and $F(t)\sim t^{-\beta_F}$
and extract the positive exponents $\beta_E$ and $\beta_F$ by fitting the
data shown in the inset of Fig.~\ref{Fig5_stressrelaxation}(b) 
and  of Fig.~\ref{Fig5_stressrelaxation}(c), respectively.
Figure~\ref{Fig6_beta-vs-gamma} 
shows the $\gamma$-dependence of $\beta_E$ (red circles)
and $\beta_F$ (blue triangles). 
The onset strain for softening $\gammas$, obtained from $\GAQS(\gamma)$, 
and the yielding threshold $\gammay$ are indicated by arrows.

As expected, we observe $\beta_E\approx 1/2$ and $\beta_F \approx 3/4$, 
in the small-$\gamma$ regime ($\gamma \lesssim 10^{-5}$) where the
linear-response approximation remains valid. 
On the other hand, 
as $\gamma$ exceeds $\gammay$, both $\beta_E$ and $\beta_F$ increase sharply and
then saturate at constant values around $\gamma=1$. 
The saturated values at the largest $\gamma$ are very close to $\beta_F=0.92$ 
and $\beta_E=2\beta_F-1=0.84$ reported previously~\cite{Chacko2019prl,Nishikawa2022prx}
for steepest-descent dynamics in two-dimensional systems 
obtained by quenching particle configurations
prepared at $T=\infty$ down to $T=0$.
These observations confirm that configurations prepared under large shear strains 
are effectively as random as those prepared at $T=\infty$ and thus share the same relaxation dynamics. 
We further find that $G(t)$ at large strain, \textit{e.g.} $\gamma =1$, decays
rapidly and does not exhibit the algebraic relaxation observed in $E(t)$. 
This can be explained by noting that, in contrast to the scalar nature of $E(t)$, 
a tensorial observable such as the stress dephases quickly and vanishes when the
initial configuration is random. 

A particularly interesting aspect is the softening regime $\gammas  < \gamma < \gammay$. 
In this range, the exponents $\beta_E=1/2$ and $\beta_F=3/4$ remain unchanged even
though the linear-response approximation 
is no longer expected to hold in this regime. 
As shown in Fig.~\ref{Fig5_stressrelaxation}(a) and (d),  
the exponents for the relaxation modulus, $\beta_G=1/2$, and for the non-affine displacement, 
$\beta_{\Dr}=1/4$, also remain identical to their linear-response values, before they
begin to deviate significantly at $\gamma > \gammay$.
In particular, the collapse of $f_G(t)$ for $\gamma < \gammay$ 
shown in the inset of
Fig.~\ref{Fig5_stressrelaxation}(a) implies a decoupling of $t$-dependence and
$\gamma$-dependence of $G(t)$, such that  
$\Delta G(t) = G(t) - G(\infty)$ can be written as
 \begin{equation}
\Delta G(t)   = \left\{G(0)-G(\infty)\right\}\ f_G(t)  
 \label{eq-wrongG}
 \end{equation}
with $f_G(t)$ which is independent of $\gamma$. 
The $\gamma$ dependence responsible for the softening is thus entirely encoded in the prefactor 
$G(0)-G(\infty)$. 
This superposition of $t^{-1/2}$ scaling and $\gamma^{1/2}$ scaling
shown in eqn~(\ref{eq-wrongG}) stands in sharp contrast to 
the frequency-domain representation, $\Delta G'(\omega, \gamma)$, given by
eqn~(\ref{eq:G_omega_gamma0}), for which the superposition picture 
breaks down at large $\gamma$ near $\gammay$
as demonstrated in Fig.~\ref{Fig1_G-vs-omega}(b).
To demonstrate how $\Delta G(t)$ in the time domain differs from that in the frequency
domain, $\Delta G(\omega, \gamma)$, we evaluate the Fourier transformation of 
eqn~(\ref{eq-wrongG}). 
The data for $f_G(t)$ is taken from the inset of
Fig.~\ref{Fig5_stressrelaxation}(a),  $G(0)$ from Fig.~\ref{Fig2_Gqs-GFT-G0} and $G(\infty)$ from  
Fig.~\ref{figA1_Gomega_GAQS}(b).  
The resulting modulus, denoted as $\Delta G'_{\mathrm{FT}}(\omega, \gamma)$ (red
open circles in Fig.~\ref{Fig2_Gqs-GFT-G0}),  shows that 
the $\gamma$-dependence resembles that of the instantaneous modulus $G(0)$.  
However, it does \textit{not} reproduce 
$\Delta G'_{\mathrm{Sim}}(\omega,\gammaz)$ obtained 
directly from oscillatory-shear simulations.
The latter shows far more pronounced softening.
 This discrepancy demonstrates that the transient relaxation dynamics is not
 equivalent to the stationary dynamics under oscillatory shear in the nonlinear
 softening regime.

\begin{figure}[tb]
    \centering
\includegraphics[width=0.8\columnwidth]{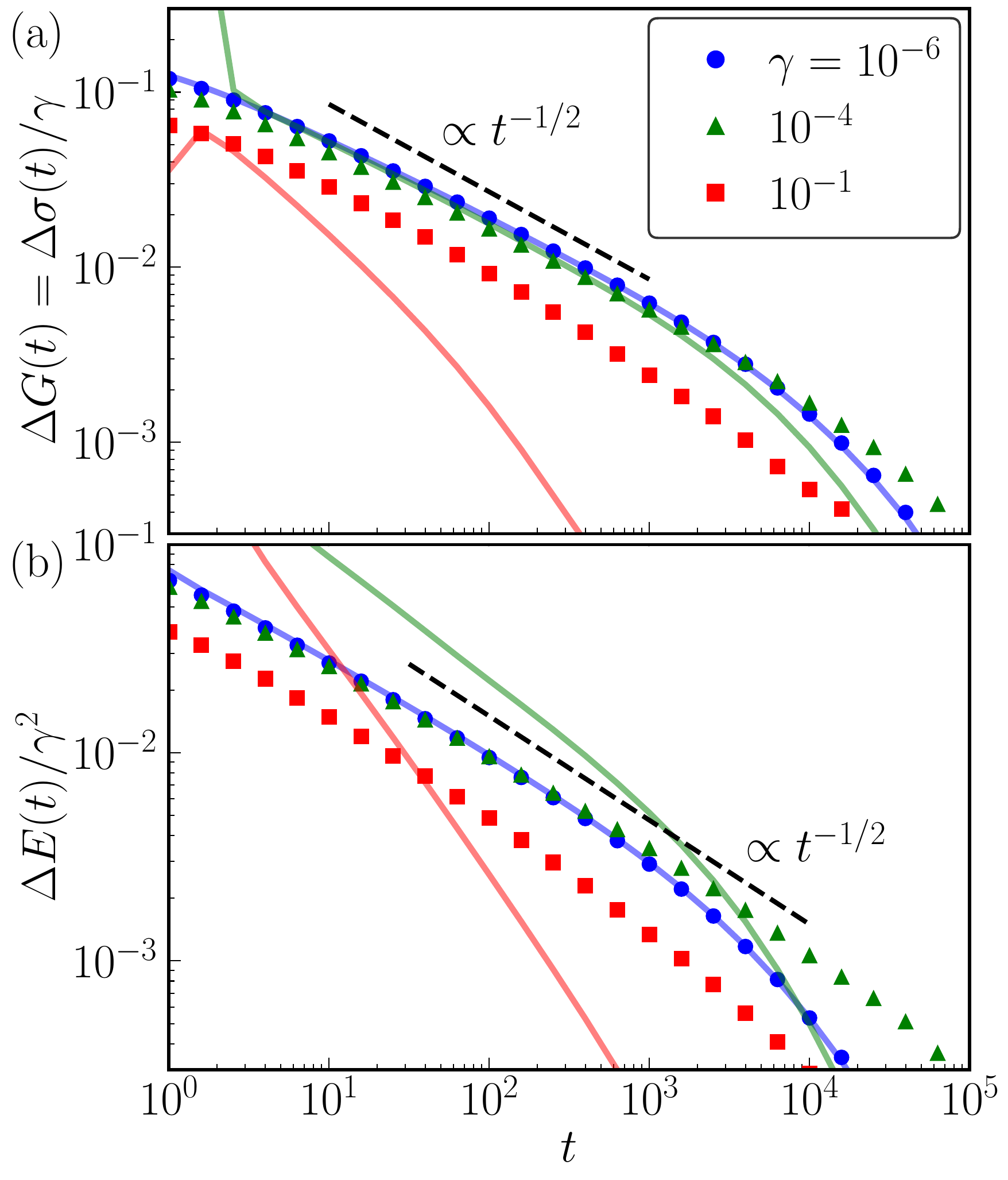}  
 \caption{(a) $\Delta G(t)$ for $\gamma=10^{-6}$ (blue), $\gamma=10^{-4}$ (green), and $\gamma=10^{-1}$ (red) for $\dphi=10^{-4}$. Dots represent the simulation data in 
Fig.~\ref{Fig5_stressrelaxation}(a) and the solid lines are the results obtained
 from the harmonic approximation (see the text).  
 The dotted  line represents $t^{-1/2}$. 
 (b) The same data for $\Delta E(t)/\gamma^2$.
 }
    \label{Fig7_vibrational_analysis} 
\end{figure}
The robustness of the exponents could in principle be attributed to the robustness of 
the harmonic approximation even beyond the linear-response regime, because  
the characteristic $t^{-1/2}$ relaxation arises if the system lies 
within a single energy basin (even if that basin is distorted by the applied shear). 
However, we find that this is not the case. 
 To show this, we compute the relaxation dynamics within the harmonic approximation 
 using the configurations of packings obtained at $t\to\infty$ after the step strain. 
 The details of the method are provided in \ref{appendixB}.
 Figure~\ref{Fig7_vibrational_analysis} shows the resulting $\Delta G(t)$ and
 $\Delta E(t)$, together with the corresponding simulation data from 
Fig.~\ref{Fig5_stressrelaxation}(a),   
 for three representative values of $\gammaz$. 
 For the smallest values $\gammaz=10^{-6}$, where the system is well within 
the linear-response regime,  the simulation and harmonic-approximation data are
essentially indistinguishable. 
For $\gammaz=10^{-4}$, which lies in the softening regime, the harmonic approximation still reproduces the $t^{-1/2}$ algebraic decay of $\Delta G(t)$, but the 
 terminal relaxation time, where the deviation from the power law begins, 
 is substantially shorter than in the simulations.
 For $\Delta E(t)$, as shown in Fig.~\ref{Fig7_vibrational_analysis}(b),
 the discrepancy between the simulation and the harmonic approximation is even more
 pronounced: 
 the approximation not only underestimates the terminal relaxation time but also
 significantly overestimates the amplitude of $\Delta E(t)$. 
 At the largest strain, $\gammaz=10^{-1}$, which is above yielding threshold, 
 the harmonic approximation fails completely to reproduce the algebraic
 relaxation, whereas the simulation data still exhibit a power-law decay,
 albeit with a slightly larger exponent.

These findings indicate that the algebraic relaxations $\Delta G(t), ~\Delta
 E(t)\sim t^{-1/2}$ are robust, but they cannot be fully explained by the harmonic approximation. 
These behaviors call for a more careful analysis of how 
the dynamics depend on the initial degree of distortion.

\section{Summary}\label{sec:sum}
In this paper, we investigated the pre-yielding mechanical response of athermal
jammed packings near the jamming transition point $\phiJ$. 
Our primary goal was to provide a unified framework for the two algebraic
scalings observed in the stress response to shear:
one is the $\omega^{1/2}$ scaling (and corresponding $t^{-1/2}$ scaling) in the
linear-response limit ($\gamma\to 0$) 
and the other is the $\gamma^{1/2}$ scaling in the quasi-static limit ($\omega \to 0$).
Motivated by the work of the Tighe group~\cite{Boschan2016sm,Dagois-Bohy2017sm} 
on the interplay between frequency and strain amplitude, we examined the two
complementary rheological protocols: 
oscillatory shear and transient relaxation.
The two protocols provide identical information in the linear-response regime,
as they are related by a simple Fourier transformation. 
However, in the nonlinear regime, they are no longer equivalent and are expected to provide complementary information. 
The oscillatory shear protocol provides detailed information on scaling behavior in stationary states at long times, whereas the roles of higher harmonics, which are assumed to be small in the present study, remain elusive and difficult to interpret. 
By contrast, the transient relaxation protocol is advantageous for directly probing real-time dynamics, where all nonlinear effects are fully taken into account.

First, using oscillatory shear, we showed that the complex modulus, 
$G^{\ast}(\omega,\gammaz)$, 
exhibits the $\omega^{1/2}$ scaling and $\gamma^{1/2}$ scaling simultaneously
as shown in eqn~(\ref{eq:G_omega_gamma0}). 
However, this behavior does not correspond to an exact 
superposition of the two scalings, as evidenced
by the quantitative differences in their onset strains.
This finding suggests a nonlinear coupling between affine and non-affine dynamics when 
the frequency is finite. 
Moreover, we found that $G^{\ast}(\omega,\gammaz)$ remains finite even below
$\phiJ$ and vanishes at a packing fraction slightly lower than $\phiJ$. 
This critical density is close to the point-to-loop reversible transition, $\phi_{\mathrm{PL}}$,
which is related to 
the reversible-irreversible (or absorbing) transition~\cite{Milz2013pre,Schreck2013pre,Nagasawa2019sm,Matsuyama2021epje,Das2010pnas}.     
This correspondence implies that the underlying geometrical transition
associated with the reversible-irreversible transition 
is closely linked to the emergence of nonlinear viscoelastic behavior near jamming.

Second, we measured the transient relaxation following a step strain.
We observed power-law relaxation, such as $t^{-1/2}$, as a real-time manifestation
of the $\omega^{1/2}$ scaling for $G(t)$ and other observables. 
This behavior is expected in the linear-response regime, where $\gamma$ is small, 
and can be rationalized within the harmonic approximation.
Remarkably, the $t^{-1/2}$ algebraic decay persists all the way up to the yielding threshold $\gammay$,
beyond which qualitatively distinct algebraic behaviors emerge.
Importantly, this persistence cannot be explained simply by the robustness of the
harmonic approximation.  

Our work serves as a starting point for understanding the nonlinear
dynamics of jammed packings. 
However, we are left with more questions than answers.
An important question concerns the connection between the mechanical response
above and below the jamming transition point $\phiJ$.  
From a geometrical standpoint, the reversible-irreversible 
transition predicts that the region where the stress remains finite corresponds
to the loop-reversible phase, in which the particles return to their original
positions after each oscillatory shear cycle, observed both above and below
$\phiJ$.  
However, the mechanical properties below $\phiJ$ differ markedly from those above. 
It would be particularly interesting to probe the distinction between the 
loop-reversible phases below and above $\phiJ$
(which is difficult to capture solely from the properties of particle trajectories)
by analyzing the relaxation dynamics of configurations across $\phiJ$. 

The robustness of the $t^{-1/2}$ scaling at large $\gamma$ is also interesting.
The study of the relaxation under shear distortion 
shares a compelling analogy with steepest descent dynamics. 
The relaxation behavior in the steepest descent dynamics sensitively 
depends on the temperature $T$ at which the initial configurations are prepared. 
If $T=\infty$, the relaxation is algebraic with the same exponents as we
reported for the strongly distorted configurations.
In the low-$T$ limit, the single basin dynamics dominates. 
In the intermediate $T$'s, the behavior of the relaxation dynamics is neither
of these limiting cases. 
We may encounter similarly hierarchical strain-dependent regimes, each 
characterized by distinct relaxation dynamics. 

From this perspective, 
the nonlinear rheology subject to frequency or time dependent perturbation
near $\phiJ$ provides a rich and fertile setting that connects several different
disciplines: the reversible-irreversible transition, the steepest-descent dynamics,
and nonlinear rheology. 
Pursuing this direction would be a worthwhile avenue for future study.

\section*{Data availability}
Data for this article are available at ZENODO at \url{https://zenodo.org/records/17746376}.

\section*{Conflicts of interest}
There are no conflicts to declare. 

\section*{Acknowledgments}

We thank M. Otsuki, A. Ikeda, H. Mizuno, H. Ikeda, Y. Nishikawa, Y. Hara,
K. Yoshii, H. Iwatsuki, and S. Tomioka for their valuable discussions. 
This work was supported by JST FOREST Program (Grant No. JPMJSP2125), JST
SPRING (Grant No. JPMJSP2125), and JSPS KAKENHI (Grant No. JP22H04472,
JP23H04503, JP23KJ1068, JP24H00192). 
H.B. would like to take this opportunity to acknowledge support from the ``THERS Make
New Standards Program for the Next Generation Researchers.''

\setcounter{section}{0}\setcounter{equation}{0}
\renewcommand{\thesection}{Appendix \Alph{section}}
\renewcommand{\theequation}{\Alph{section}\arabic{equation}}
\section{FIRE algorithm and shear-stabilization protocol}\label{appendixA0}
In this section, we describe the details of the FIRE algorithm and the shear-stabilization protocol used in this study.

The FIRE algorithm is employed for all energy-minimization procedures, including
the preparation of initial configurations via compression–decompression cycles,
as well as quasi-static simple shear. 
Within this algorithm, we perform molecular dynamics simulations using Newton's equations of motion,
\begin{equation}
    m\ddot{\bm{r}}_j=-\frac{\partial U}{\partial\bm{r}_j}.
    \label{eq:eom_FIRE}
\end{equation}
We integrate this equation using a semi-implicit Euler scheme.
The time-step width is chosen in the range $\Delta t \in [10^{-6},0.05]$.
At each step, the total power is given by
\begin{equation}
    p=-\sum_{j=1}^N\dot{\bm{r}}_j\cdot\frac{\partial U}{\partial\bm{r}_j},
\label{eq:power_FIRE}
\end{equation}
and the particle velocities are updated as
\begin{equation}
    \dot{\bm{r}}_j=(1-\alpha)\dot{\bm{r}}_j+\alpha\hat{\bm{F}}_j|\dot{\bm{r}}_j|,
\label{eq:v_update}
\end{equation}
where $\hat{\bm{F}}_j$ is the unit vector in the direction of $-\partial U/\partial\bm{r}_j$.
If $p>0$ for five consecutive steps, $\Delta t$ and $\alpha$ are updated to $\min(1.1\Delta t,0.05)$ and $0.99\alpha$, respectively.
If $p\leq 0$, $\Delta t$ and $\alpha$ are reset to $0.5\Delta t$ and $0.1$, respectively.
These parameters are identical to those used in Ref.~\cite{Kawasaki2024prl}.
The procedure is then repeated by returning to eqn~(\ref{eq:eom_FIRE}) and recalculating eqns.~(\ref{eq:power_FIRE}) and (\ref{eq:v_update}).

We incorporate a shear-stabilization protocol into the FIRE algorithm during the compression–decompression process.
In addition to the particle equations of motion, we solve the equation of motion for $\gamma$,
\begin{equation}
    \ddot{\gamma}=-\frac{\sigma}{W},
\end{equation}
where $W=(100L/\sigma_{\mathrm{S}})^{-1}$, as proposed in Ref.~\cite{Kawasaki2024prl}.
Analogous to eqn~(\ref{eq:v_update}), the shear rate is updated as
\begin{equation}
\dot{\gamma}=(1-\alpha)\dot{\gamma}+\alpha\frac{\sigma}{|\sigma|}|\dot{\gamma}|,
\end{equation}
where $\sigma/|\sigma|$ denotes the sign of the shear stress.
We use the same definition of the power $p$ as given in eqn~(\ref{eq:power_FIRE}), while neglecting the contribution from $\sigma\dot{\gamma}$.

\section{Supplemental data for oscillatory shear}\label{appendixA}

\begin{figure}[tb]
  \centering
  \includegraphics[width=0.8\columnwidth]{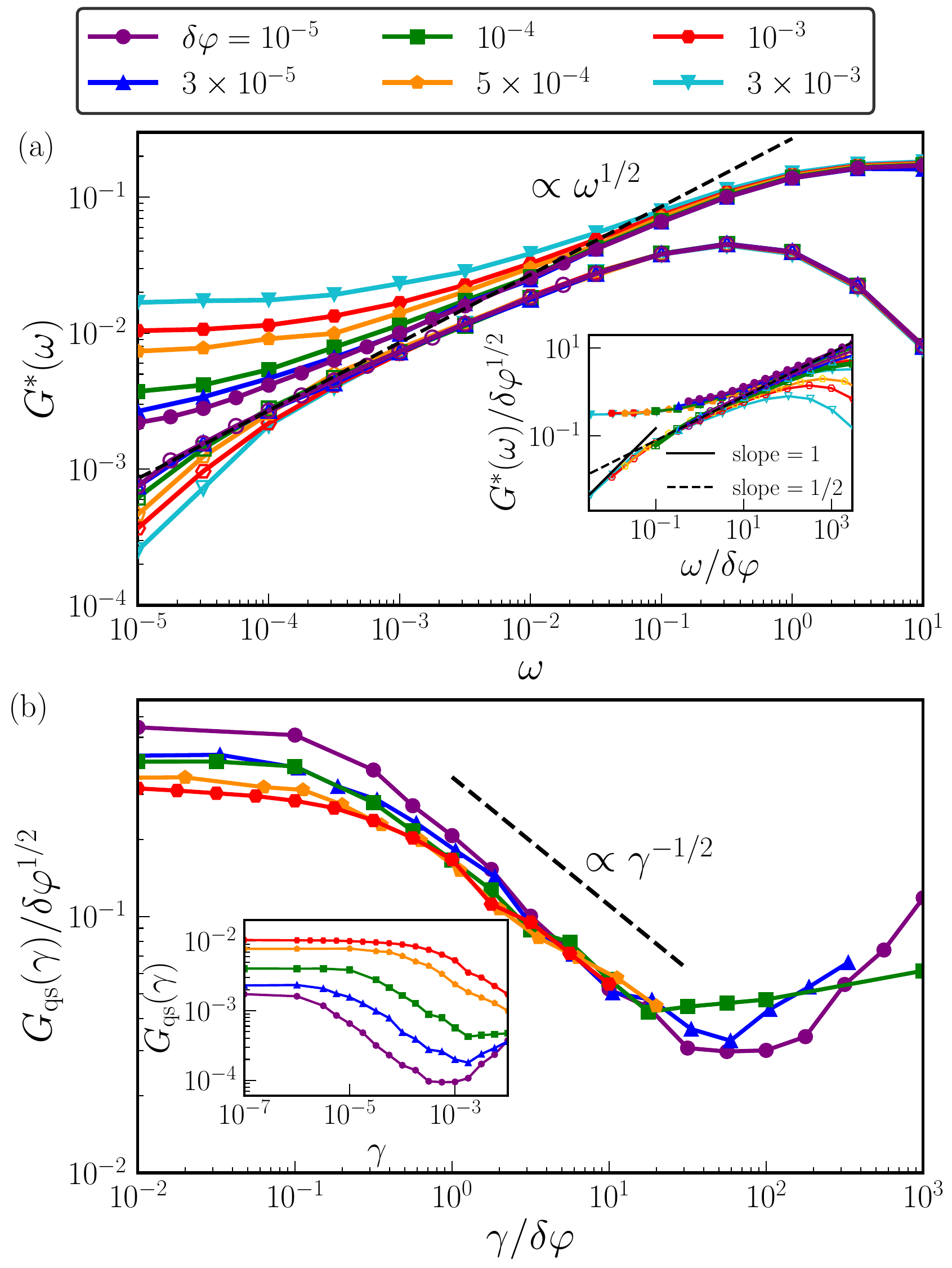}
  \caption{(a) $\omega$ dependence of $G'(\omega)$ (the filled symbols) and $G''(\omega)$ (the open symbols) for various $\dphi$ at $\gamma=10^{-7}$. The dashed line represents $\omega^{1/2}$. 
  (Inset) $G^*(\omega)/\dphi^{1/2}$ vs $\omega/\dphi$. 
  The black solid and dashed lines represent $G''(\omega)\propto\omega$ and
 $G'(\omega), G''(\omega)\propto\omega^{1/2}$, respectively. 
  (b) $\GAQS(\gamma)/\dphi^{1/2}$ as a function of $\gamma/\dphi$ for various $\dphi$. The dashed line indicates $\GAQS(\gamma)\propto\gamma^{-1/2}$. (Inset) The raw data of $\GAQS(\gamma)$.}
  \label{figA1_Gomega_GAQS}
\end{figure}
We first reproduce the $\omega$-dependence of the complex modulus, $G^*(\omega)=G'(\omega)+iG''(\omega)$, in the linear-response regime.
Figure~\ref{figA1_Gomega_GAQS}~(a) shows the $\omega$-dependence 
of the storage modulus, $G'(\omega)$, and the loss modulus, $G''(\omega)$, for
several values of $\delta\varphi$ at $\gammaz=10^{-7}$, well below the onset of
the softening regime $\gammas$. 
The qualitative features of the $\omega$-dependence are identical to those
reported in previous studies~\cite{Tighe2011prl,Baumgarten2017sm,Hara2025natp}: 
for $\omega>\omega_{\mathrm{s}}$, both $G'(\omega)$ and $G''(\omega)$ exhibit 
$\omega^{1/2}$ power-law behavior.
Note that the $\omega^{1/2}$-scaling regime for $G''(\omega)$ extends to smaller $\omega$ 
than $\omega_{\mathrm{s}}$, as reported by Hara \textit{et al.}~\cite{Hara2025natp}.
For $\omega\ll\omega_{\mathrm{s}}$, $G'(\omega)$ converges to its long-time
limit $G(t\to\infty)$, whereas $G''(\omega)$ is proportional to $\omega$.
At high frequencies, $\omega\gtrsim1$, $G'(\omega)$ saturates to the short-time
limit of the relaxation modulus, $G(t=0)$ (see
Sec.~\ref{sec:stress_relaxation}), whereas $G''(\omega)$ starts to decrease as reported in Ref.~\cite{Baumgarten2017sm}, 
which originates from the microscopic dynamics.
In this study, we used Stokes drag in the equation of motion for the particles (eqn~(\ref{eq:EoM_shear})), 
which is suitable for describing non-Brownian particles), instead of the Durian drag
model used in other studies~\cite{Tighe2011prl,Baumgarten2017sm,Dagois-Bohy2017sm}. 
A noticeable feature of Fig.~\ref{figA1_Gomega_GAQS}(a) is that the window over
which the $\omega^{1/2}$ scaling for $G''(\omega)$ holds is relatively narrow
compared with that of $G'(\omega)$, owing to the crossover to the high-frequency regime. 
This limited scaling range makes the identification of the
$\omega^{1/2}$ scaling in our study more challenging, as discussed in Sec.~\ref{sec:oscillatory}.

In Figure~\ref{figA1_Gomega_GAQS}(b), we show the $\gammaz$-dependence of 
the modulus in the quasi-static limit, $\GAQS=G'(\omega\to0, \gamma)$, 
which reproduces the results 
of the previous studies~\cite{Otsuki2014,Boschan2016sm,Otsuki2022prl,Kawasaki2024prl}.
We observe that $\GAQS(\gamma)$ exhibits softening as $\GAQS(\gamma)\sim\gammaz^{-1/2}$ 
at $\gammaz>\gammas$, and then saturates and eventually increases as the system
enters the hardening regime at large $\gammaz$~\cite{Kawasaki2024prl}.

\begin{figure}[tb]
    \centering
    \includegraphics[width=0.95\columnwidth]{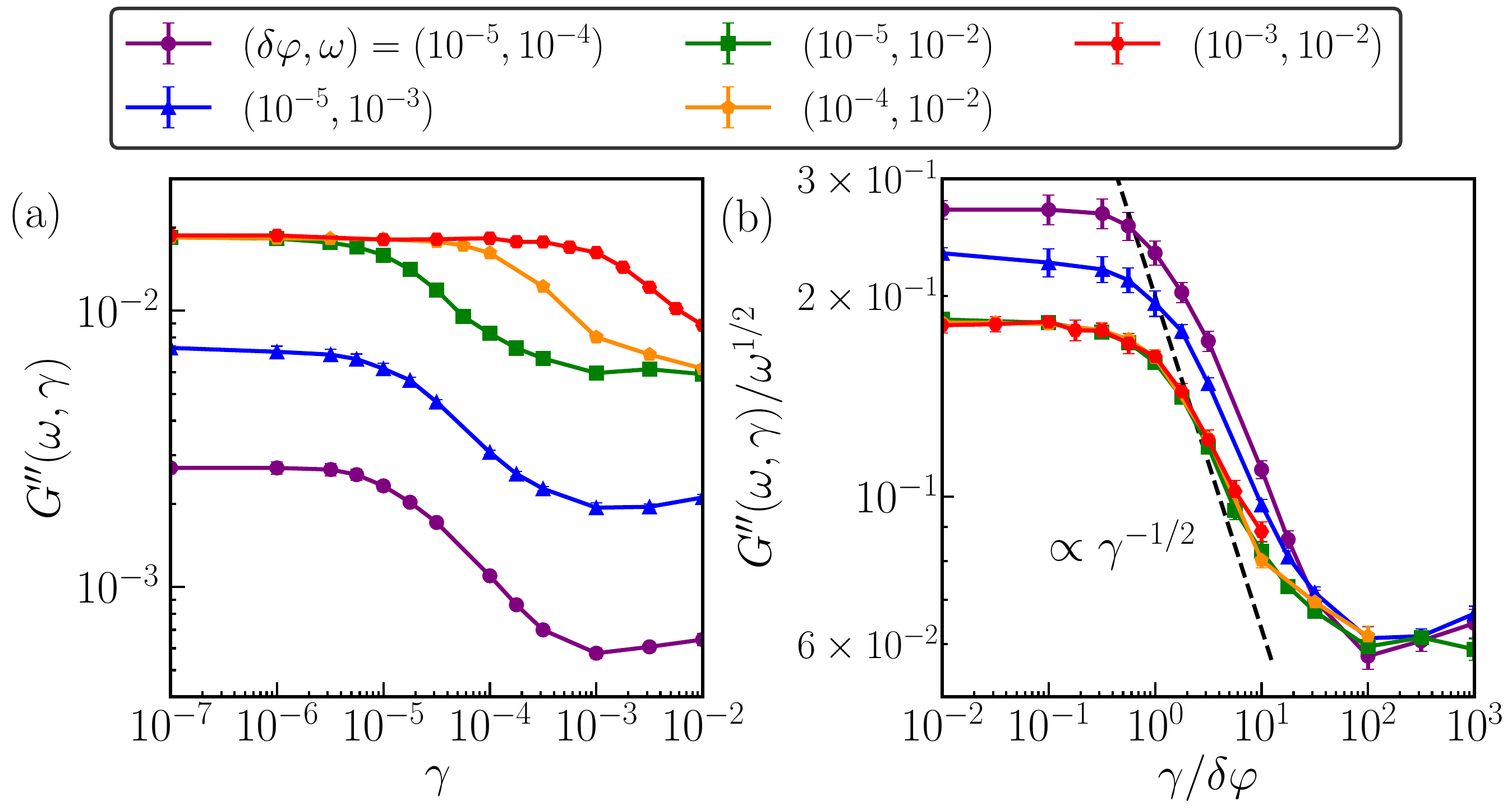}
    \caption{
    (a) $\gamma$-dependence of $G''(\omega,\gamma)$ for several sets of $\dphi$ and $\omega$. 
    (b) Scaled $G''(\omega, \gamma)/\omega^{1/2}$ plotted as a function of $\gamma/\dphi$. 
    The error bars in all panels represent the standard error of the mean.
}
 \label{FigA2_G2-vs-gamma}
\end{figure}
Figure~\ref{FigA2_G2-vs-gamma} shows the $\gammaz$-dependence of the loss
modulus $G''(\omega,\gammaz)$, obtained by using the same set of parameters as those for
$\Delta G'(\omega, \gamma)$ shown in Fig.~\ref{Fig1_G-vs-omega}. 
The $\gammaz$-dependence of $G''(\omega,\gammaz)$ is qualitatively similar to
that of $\Delta G'(\omega, \gamma)$, but slight quantitative differences are observed.
The shear-softening exponent is smaller than $1/2$,
as shown in Fig.~\ref{FigA2_G2-vs-gamma}~(b). 
This discrepancy originates from the narrow $\omega^{1/2}$-scaling window 
already observed in the linear-response regime (Fig.~\ref{figA1_Gomega_GAQS}(a)).

\begin{figure}[tb]
    \centering
    \includegraphics[width=0.8\columnwidth]{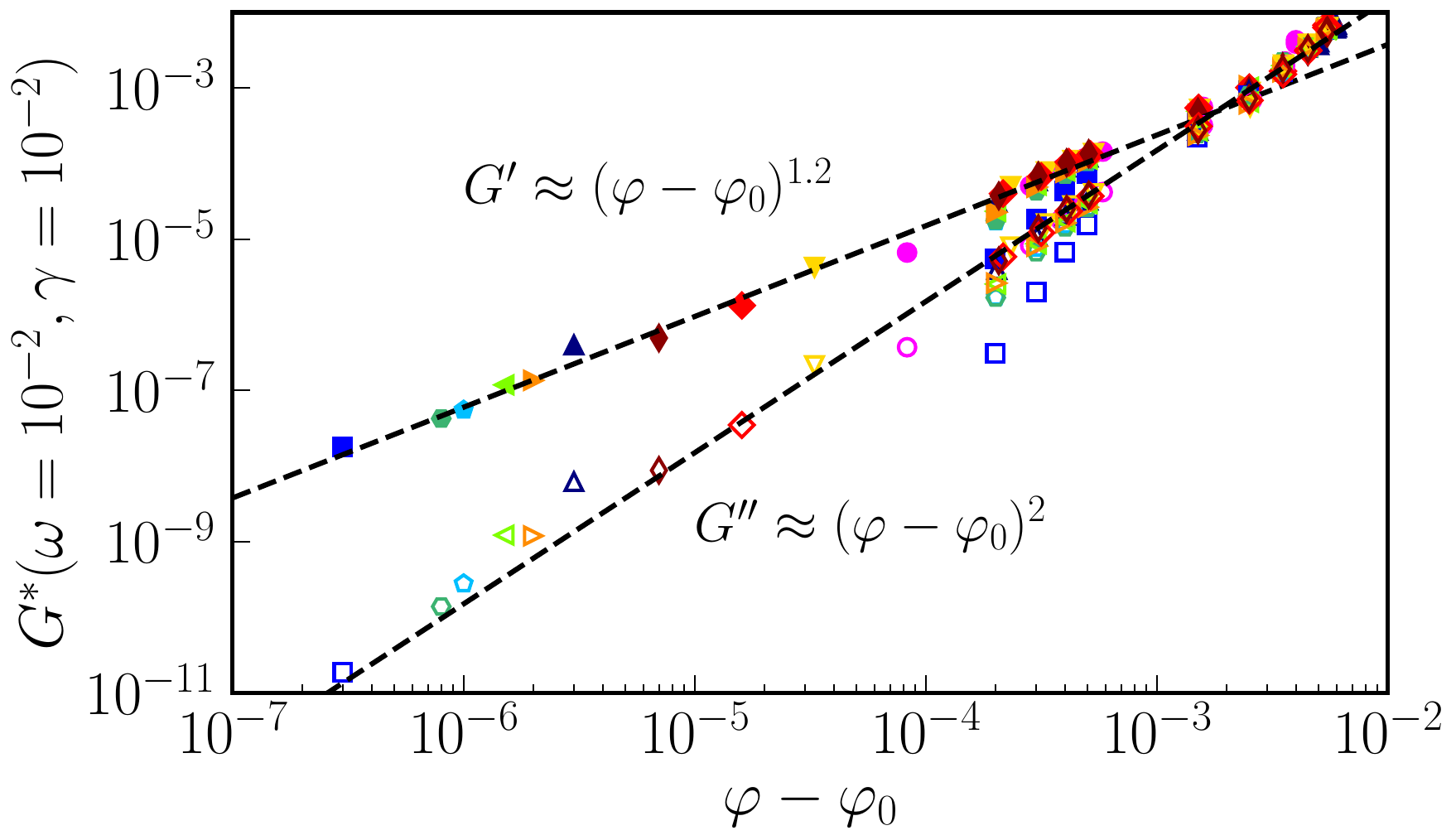}
    \caption{
    $G'(\omega, \gamma)$ (filled symbols) and $G''(\omega, \gamma)$ (open
 symbols)  as a function of $\phi-\phi_{0}$ at $\omega=10^{-2}$ and
 $\gamma=10^{-2}$.  
    The data in Fig.~\ref{Fig3_3d-G} are used. 
    Different symbols correspond to different samples. 
    Dashed lines represent $(\phi-\phi_0)^{\beta}$ with $\beta =1.2$ for
 $G'(\omega,\gamma)$ and $\beta=2$ for $G''(\omega,\gamma)$.  
}
\label{FigA3_G-below-phiJ}
\end{figure}
Finally, we examine how the complex modulus $G^{\ast}(\omega, \gamma)$ vanishes as 
$\varphi$ is decreased.   
As shown in Fig.~\ref{Fig3_3d-G}, $G'(\omega, \gamma)$ decreases to zero below $\phiJ$, 
and the density at which it vanishes, $\varphi_{0}$, decreases with increasing $\gammaz$. 
We plot $G'(\omega, \gamma)$ and $G''(\omega, \gamma)$ as a function of
$\varphi-\varphi_{0}$ and fit them with a power law, 
$G'(\omega, \gamma),\;G''(\omega, \gamma)\propto(\varphi-\varphi_{0})^{\beta}$,  
where $\varphi_{0}$ and $\beta$ are fitting parameters.  
Figure~\ref{FigA3_G-below-phiJ} shows the results of the fitting at 
$\omega=10^{-2}$ and $\gammaz=10^{-2}$.  
Note that, similar to $\phiJ$, $\varphi_{0}$ varies from sample to sample; therefore, we performed the fitting using the individual $\varphi_{0}$ for each configuration.
We find that the exponent 
$\beta$ for $G'(\omega, \gammaz)$ ($\beta\approx1.2$)
and for $G''(\omega,\gammaz)$ ($\beta\approx2$) differ.

\section{Supplemental data for transient stress relaxation}\label{appendixB}\setcounter{equation}{0}

We briefly review the derivation of the power-law relaxation 
for the shear stress and energy, 
$\sigma(t), E(t) \sim t^{-1/2}$, using the 
harmonic approximation based on the vibrational mode analysis~\cite{Nishikawa2022prx}.  
We consider the dynamics of the shear stress in terms of the displacement vector
$\bm{u}(t)$ and the Hessian matrix $\mathcal{H}$ 
of the inherent structure of a packing. 
The potential energy can be expressed as
\begin{equation}
E(t)\simeq E(t\to\infty)+\frac{1}{2}\bm{u}^{\mathrm{T}}(t)\mathcal{H}\bm{u}(t),
    \label{eq:harmonic_approx}
\end{equation}
where $\mathcal{H}$ is the Hessian matrix and 
$\bm{u}(t) = \bm{r}(t)- \bm{r}^{\mathrm{eq}}$ 
is the $dN$-dimensional displacement vector of the $N$ particles from their 
mechanical equilibrium position $\bm{r}^{\mathrm{eq}} = \bm{r}(t \to\infty)$,
with $d$ being the spatial dimension.  
The shear stress can be expanded in $\bm{u}(t)$ as
\begin{equation}
\sigma(\bm{r}^{\mathrm{eq}} + \bm{u}(t)) \approx
 \sigma(\bm{r}^{\mathrm{eq}}) + \bm{\Xi}_{xy}(\bm{r}^{\mathrm{eq}}) \cdot \bm{u}(t),  
\end{equation}
where $\bm{\Xi}_{xy}$ is the shear stress gradient also known as the affine force field defined by
\begin{equation}
    \bm{\Xi}_{xy}(\bm{r}^{\mathrm{eq}}) = \left( \frac{\partial \sigma(\bm{r}^{\mathrm{eq}}_1)}{\partial \bm{r}_1}^{\mathrm{T}}, \cdots, \frac{\partial \sigma(\bm{r}^{\mathrm{eq}}_N)}{\partial \bm{r}_N}^{\mathrm{T}} \right)^{\mathrm{T}}.
\end{equation}
For the virial shear stress given in eqn~(\ref{eq:virial_stress}), the gradient term with respect to particle $j$ is
\begin{equation}
    \bm{\Xi}_{xy}^j = -\frac{1}{V} \sum_{k \neq j} (\kappa_{jk} r_{jk} - t_{jk})
     n_{jk}^x n_{jk}^y \bm{n}_{jk}, 
\end{equation}
where $\kappa_{jk} = \dd^2U / \dd r_{jk}^2$, $t_{jk} = \dd U /\dd r_{jk}$, and
$\bm{n}_{jk} = \bm{r}_{jk} / r_{jk}$.  
Using the eigenvalues $\{\lambda_k\}$ and corresponding eigenvectors $\{\bm{e}_k\}$
of the Hessian matrix, the displacement vector can be expanded as 
$\bm{u}(t) = \sum_k c_k(t)\bm{e}_k$,  
with the amplitude given by $c_k(t) = \bm{u}(t) \cdot \bm{e}_k$
due to the orthogonality of $\bm{e}_k$. 
Under the harmonic approximation, the dynamics follow the linearized equation of
motion $-\zeta\dot{\bm{u}}(t)=\mathcal{H} \bm{u}(t)$, 
which gives the time evolution of $c_k(t)$ as 
$c_k(t) = c_k(0) e^{-\lambda_k t/\zeta}$.
Using the relationship,
$c_k(t) = \bm{u}(t) \cdot \bm{e}_k$, mentioned above, 
the relaxation of the shear stress, 
$\Delta \sigma(t) =\sigma(\bm{r}^{\mathrm{eq}} +
\bm{u}(t))-\sigma(\bm{r}^{\mathrm{eq}})$,  
can be expressed as
\begin{equation}
\Delta \sigma(t) =\sum_{k} \left(\bm{\Xi}_{xy}(\bm{r}^{\mathrm{eq}}) \cdot \bm{e}_k\right)
c_k(0) e^{-\lambda_k t/\zeta} 
=\int   \dd\lambda\,\rho(\lambda)A(\lambda)e^{-\lambda t/\zeta}, 
\label{eq:Delta_sigma}
\end{equation}
where $\rho(\lambda)=(dN-d)^{-1}\sum_k\delta(\lambda-\lambda_k)$ is the density
of eigenvalues, and 
$A(\lambda)=(\bm{\Xi}_{xy}(\bm{r}^{\mathrm{eq}})
\cdot\bm{e})(\bm{u}(0)\cdot\bm{e})$  
represents the weight of each mode for the stress relaxation.    
The eigenvalue density is related to the vibrational density of states $D(\Omega)$ 
by $\rho(\lambda)=D(\Omega)(\dd\Omega/\dd\lambda)$, where $\Omega = \sqrt{\lambda}$.

Similarly, the energy relaxation can be derived by expanding the energy using $\bm{u}(t)$ and
$\mathcal{H}$ (see eqn~(\ref{eq:harmonic_approx})).  
With the linearized equation of motion described above, the energy
relaxation follows: 
\begin{equation}
    \Delta E(t) = \frac{1}{2}\sum_k c_k^2(0) \lambda_k e^{-2\lambda_k t/\zeta}
     =\frac{1}{2}\int \dd\lambda\,
     \rho(\lambda)B(\lambda)\lambda e^{-2\lambda   t/\zeta}, 
    \label{eq:Delta_E}
\end{equation}
where $B(\lambda)=(\bm{u}(0)\cdot\bm{e})^2$ is 
the weight of each mode for the energy relaxation~\cite{Nishikawa2022prx}. 

In the vicinity of $\phiJ$, $D(\Omega)$ develops a plateau at low frequencies,
so that $\rho(\lambda)\sim \lambda^{-1/2}$.
As shown below, $A(\lambda)$ in eqn~(\ref{eq:Delta_sigma}) is approximately constant, whereas 
$B \sim \lambda^{-1}$ in eqn~(\ref{eq:Delta_E}).  
Therefore, assuming these integrals are dominated by the low-$\lambda$ contribution, both 
$\Delta \sigma(t)$ and $\Delta E(t)$ 
scale as
$t^{-1/2}$, which reproduces the results of Figure~\ref{Fig4_linear_stressrelaxation} in the linear-response regime. 

\begin{figure}[tb]
    \centering
    \includegraphics[width=0.8\columnwidth]{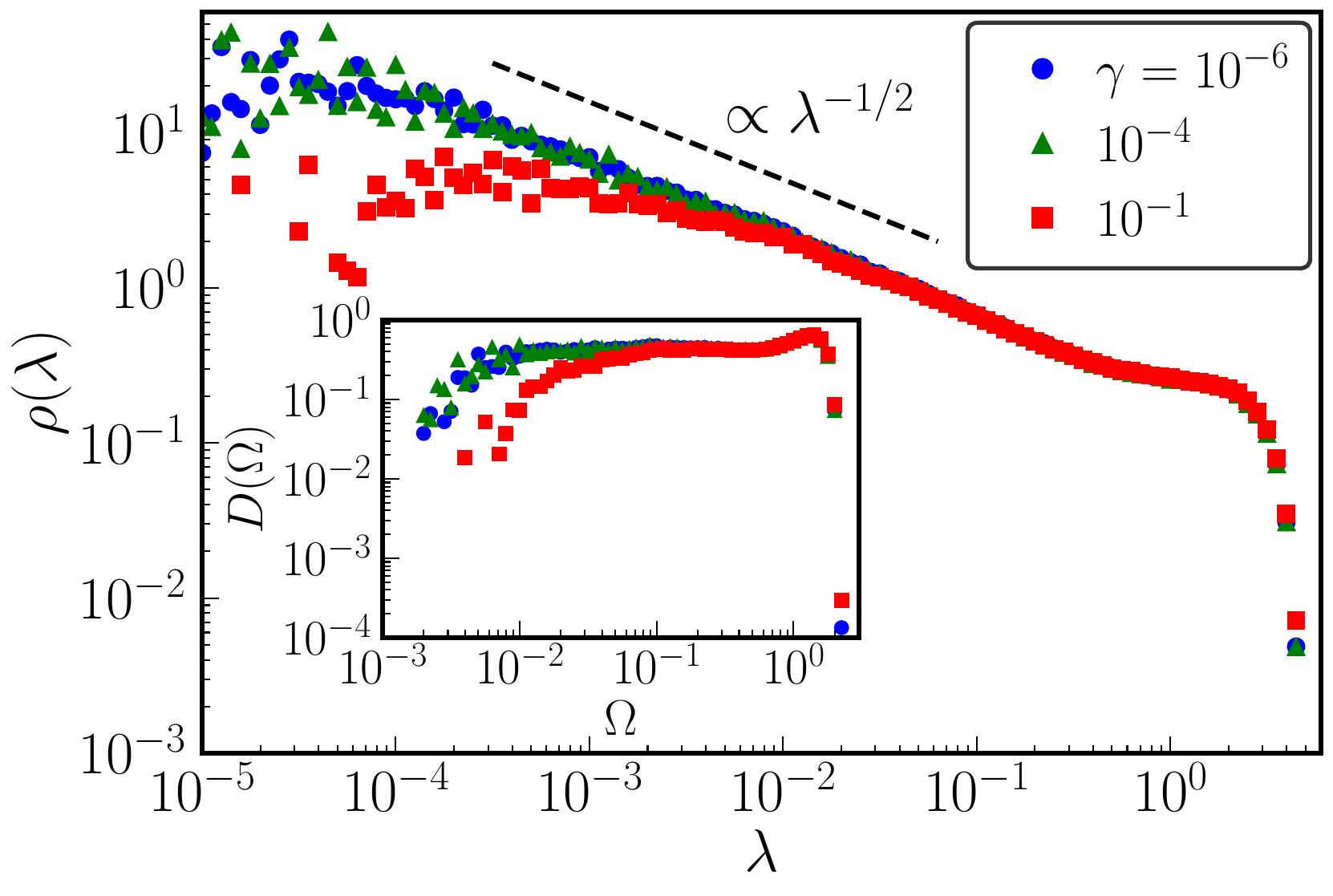}
    \caption{The density of eigenvalues $\rho(\lambda)$ in mechanical
 equilibrium for $\gamma=10^{-6}$ (blue), $10^{-4}$ (green), and $10^{-1}$
 (red) at $\dphi = 10^{-4}$.  
    The dashed line indicates $\lambda^{-1/2}$. (Inset) the vibrational density of states $D(\Omega)$.}
    \label{FigB1_vDoE}
\end{figure}
\begin{figure}[tb]
    \centering
    \includegraphics[width=0.95\columnwidth]{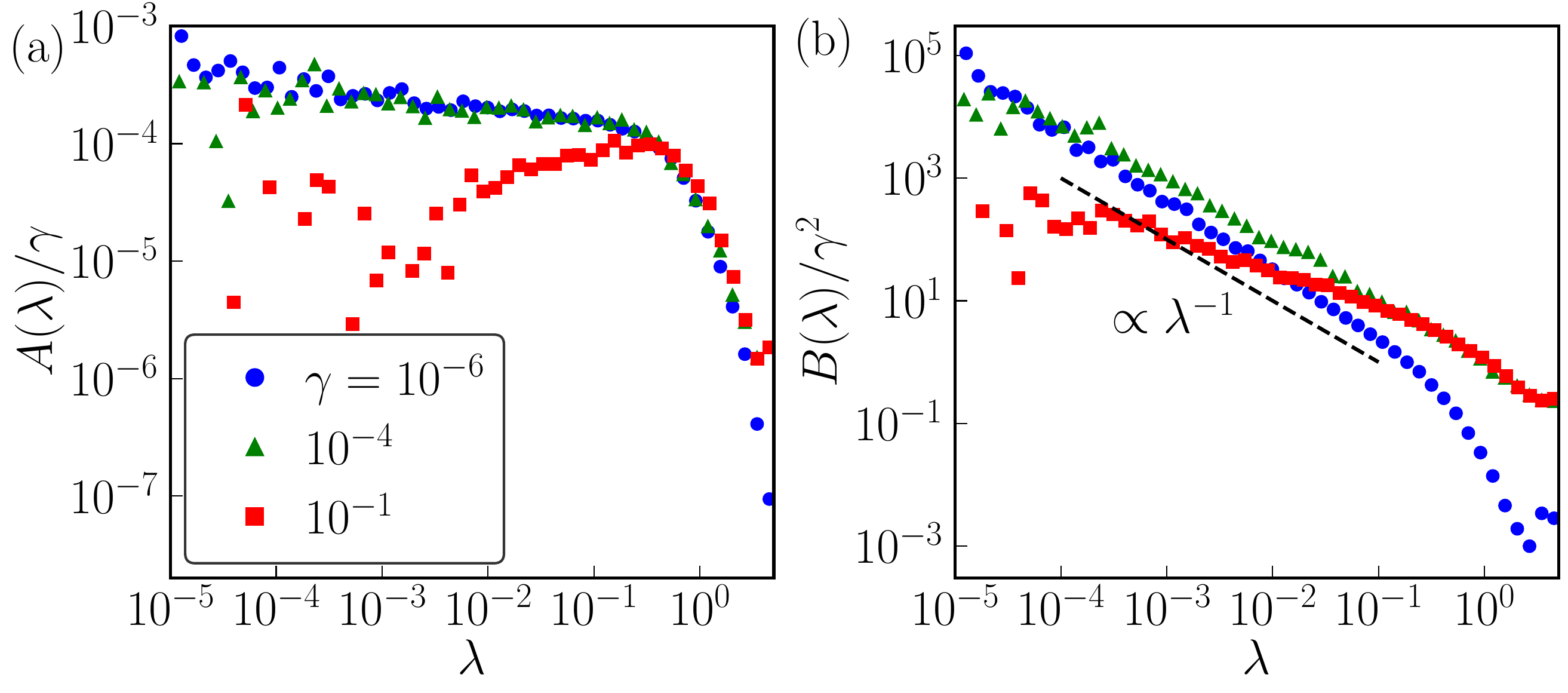}
    \caption{
$\lambda$-dependence of (a) $A(\lambda) = (\bm{\Xi}_{xy}(\bm{r}^{\mathrm{eq}})
 \cdot 
 \bm{e})(\bm{u}(0)\cdot\bm{e})$ and (b) $B(\lambda) = (\bm{u}(0) \cdot
 \bm{e})^2$ scaled by $\gamma$ for $\gamma=10^{-6}$ (blue), $\gamma=10^{-4}$
 (green) , and $\gamma=10^{-1}$ (red) at $\dphi=10^{-4}$. 
    The dashed lines in panel (b) indicate $\lambda^{-1}$ for $B(\lambda)$.
}
    \label{FigB2_AB}
\end{figure}
Next, we perform the harmonic approximation analysis for sheared systems.  
We directly compute the eigenvalues of the Hessian matrix $\mathcal{H}$ from 
numerically obtained jammed packings.  
Figure~\ref{FigB1_vDoE} shows the density of eigenvalues
$\rho(\lambda)$ and the vibrational density of states $D(\Omega)$ (inset) for
the packing at $\gammaz=10^{-6}$, $10^{-4}$, and $10^{-1}$, at $\delta\varphi = 10^{-4}$.   
$D(\Omega)$ (and hence $\rho(\lambda)$) for the first two values of $\gammaz$ 
are almost identical to the unstrained system, except for a small deviation at low-$\Omega$.
They exhibit a plateau, a feature universally observed near $\phiJ$~\cite{Silbert2005prl,Silbert2009pre}.  
$D(\Omega)$ for $\gammaz=10^{-1}$ is also similar except for a larger onset frequency of the plateau, $\Omega_{\ast}$, 
which is due to the decrease of $\phiJ$ at large $\gammaz$~\cite{Kawasaki2024prl}.
Using the computed eigenvalues and eigenvectors shown in
Fig.~\ref{FigB1_vDoE}, 
we evaluate $A(\lambda)$ and
$B(\lambda)$ from eqns~(\ref{eq:Delta_sigma}) and (\ref{eq:Delta_E}).   
Figures~\ref{FigB2_AB}(a) and (b) show the $\lambda$-dependence 
of $A(\lambda)$ and $B(\lambda)$.  
For $\gammaz=10^{-6}$ and $10^{-4}$, we find that 
$A(\lambda) \sim \text{const.}$ and 
$B(\lambda) \sim \lambda^{-1}$ over a broad range of $\lambda$, 
corresponding to the plateau of $D(\Omega)$.
The small difference of $B(\lambda)$ for the two values of $\gammaz$ is
reflected in the overestimation of $\Delta E(t)$ in the softening regime, as
shown in Fig.~\ref{Fig7_vibrational_analysis}.
For $\gammaz=10^{-1}$, on the other hand, both $A(\lambda)$ and $B(\lambda)$ differ 
substantially from those for other $\gammaz$. 

\providecommand*{\mcitethebibliography}{\thebibliography}
\csname @ifundefined\endcsname{endmcitethebibliography}
{\let\endmcitethebibliography\endthebibliography}{}

\end{document}